\documentclass[twocolumn,twocolappendix]{aastex631}

\newcommand{\onepass}{\texttt{hst1pass}}
\newcommand{\kstwo}{\texttt{ks2}}
\newcommand{\collate}{\texttt{hst2collate}}

\begin{document}

\title{A re-analysis of the isolated black hole candidate OGLE-2011-BLG-0462/MOA-2011-BLG-191}

\author[0000-0002-6406-1924]{Casey Y. Lam}
\correspondingauthor{Casey Y. Lam} 
\email{clam@carnegiescience.edu}
\affiliation{University of California, Berkeley, Department of Astronomy, Berkeley, CA 94720}

\author[0000-0001-9611-0009]{Jessica R. Lu}
\affiliation{University of California, Berkeley, Department of Astronomy, Berkeley, CA 94720}

\begin{abstract}

There are expected to be $\sim 10^8$ isolated black holes (BHs) in the Milky Way.
OGLE-2011-BLG-0462/MOA-2011-BLG-191 (OB110462) is the only such BH with a mass measurement to date.
However, its mass is disputed: \citet{Lam:2022a,Lam:2022b} measured a lower mass of $1.6 - 4.4 M_\odot$, while \citet{Sahu:2022, Mroz:2022} measured a higher mass of $5.8 - 8.7 M_\odot$.
We re-analyze OB110462, including new data from the Hubble Space Telescope (HST) and re-reduced Optical Gravitational Lensing Experiment (OGLE) photometry. 
We also re-reduce and re-analyze the HST dataset with newly available software.
We find significantly different ($\sim 1$ mas) HST astrometry than \cite{Lam:2022a,Lam:2022b} in the de-magnified epochs due to the amount of positional bias induced by a bright star $\sim$0.4 arcsec from OB110462.
After modeling the updated photometric and astrometric datasets, we find the lens of OB110462 is a $6.0^{+1.2}_{-1.0} M_\odot$ BH.
Future observations with the Nancy Grace Roman Space Telescope, which will have an astrometric precision comparable or better to HST but a field of view $100\times$ larger, will be able to measure hundreds of isolated BH masses via microlensing. 
This will enable the measurement of the BH mass distribution and improve understanding of massive stellar evolution and BH formation channels. 

\end{abstract}

\section{Introduction \label{sec:Introduction}}

Although massive stars, the progenitors of black holes (BHs), are typically born in binaries, the majority of the Milky Way’s $10^7 - 10^9$ BHs are expected to be isolated \citep{Fender:2013, Wiktorowicz:2019, Olejak:2020}.
Around 20-30\% of O-stars are expected to merge and form a single, even more massive star \citep{Sana:2012}, and many of the remaining binary systems are disrupted before, during, or after the formation of the BH due to natal kicks or mass loss.
Despite this, nearly all known Galactic BHs are in binary systems \citep{Corral-Santana:2016, Thompson:2019, El-Badry:2023_sun, Chakrabarti:2022, El-Badry:2023_rg}.
This detection bias exists because, unlike BH binaries, isolated BHs do not have a companion that can electromagnetically identify their presence, making them particularly elusive.
Detecting and characterizing isolated BHs is a critical first step needed to understand the full Galactic BH population.

Gravitational lensing is the most practical way to detect isolated BHs, as the observational signature depends only on the mass of the lens, and not its luminosity. 
In particular, \emph{microlensing}, the regime of gravitational lensing where the images are unresolved, provides a way to find and measure the masses of dark objects.
As a foreground lens (e.g. a BH) aligns in front of a background source of light (e.g. a Bulge star), this causes the background source to temporarily brighten and change its apparent position; the transient brightening is called \emph{photometric microlensing} and the transient change in position is called \emph{astrometric microlensing}.
The combination of the photometric and astrometric signals can be used to measure the mass, distance, and proper motion of the lens \citep{Hog:1995, Miyamoto:1995, Walker:1995}.
For more details on astrometric microlensing, see \cite{Dominik:2000}.

\subsection{An isolated dark compact object found with microlensing \label{sec:An isolated dark compact object found with microlensing}}

OGLE-2011-BLG-0462/MOA-2011-BLG-191 (hereafter OB110462) is the first isolated, dark compact object to have its mass measured with astrometric microlensing.
It was identified as a microlensing event toward the Galactic Bulge located at (17:51:40.19, -29:53:26.3) and has been observed both photometrically and astrometrically in order to measure the lens' mass.
However, the nature of OB110462's lens is disputed.
\cite{Sahu:2022} inferred the lens to be a $M_L = 7.1 \pm 1.3 M_\odot$ dark object, making OB110462 a firm BH detection similar in mass to other known Galactic BHs in binary systems.
\cite{Lam:2022a,Lam:2022b} inferred a lower mass object, which depending on the modeling, led to a $M_L = 2.15^{+0.67}_{-0.54} M_\odot$ or $M_L = 3.79^{+0.62}_{-0.57} M_\odot$ dark object, implying a neutron star or low-mass BH.
Both groups analyzed slightly different subsets of high-cadence ground-based photometry and high-resolution Hubble Space Telescope (HST) astrometry; see \cite{Mroz:2022} for a summary. 
In particular, despite analyzing the same astrometric data, both groups derived different stellar positions.
In addition, they both found the microlensing parameters inferred from the ground-based photometry were in tension with the parameters inferred by their respective astrometric measurements.

\cite{Mroz:2022} re-analyzed the ground-based OGLE photometry of OB110462. 
They found systematics in the photometry resulting from imperfect image subtraction due to variations in seeing.
Updated modeling using the revised data showed that the OGLE photometry could be self-consistently modeled with the astrometry of \cite{Sahu:2022}, and \cite{Mroz:2022} inferred a lens mass of $M_L = 7.88 \pm 0.82 M_\odot$.

\cite{Mereghetti:2022} combined new and archival Chandra imaging to search for X-ray emission from OB110462.
No X-rays were detected, and based on the detection upper limits, concluded OB110462 could be consistent with an accreting isolated BH with low radiative efficiency.
A neutron star moving slowly or in a high density environment would be disfavored, but uncertainties in the velocity and environment density, as well as the accretion efficiency, did not allow stronger statements to be made on the nature of OB110462.

\subsection{Rationale for re-analysis of OB110462}
Since publication of the initial discovery and modeling papers of OB110462 in July 2022, there have been several new developments.
First, as mentioned in \S \ref{sec:An isolated dark compact object found with microlensing}, there are updated ground-based photometry data from OGLE.
There are also two additional HST data points for OB110462.
OB110462 was one of 70 targets in a HST snapshot program to image microlensing events (SNAP-16716; PI: K. Sahu); observations of OB110462 were taken in May 2022.
In addition, the second and final epoch of a Cycle 29 program (GO-16760; PI: C. Lam) to obtain OB110462 astrometry was taken in September 2022.
With regards to analysis tools, an updated version of the software used in the extraction of the astrometry from the HST data \citep[\onepass,][]{Anderson:2022} was released in July 2022.

The ability to find and characterize isolated BHs is necessary to understand the evolution and death of massive stars. 
In turn, massive stars impact our understanding of a wide range of astrophysical problems, from the high-mass end of the stellar initial mass function, to chemical evolution, to galactic feedback.
Without understanding the properties of isolated BHs, these problems cannot be solved.
Thus, a re-analysis of OB110462 is a timely and worthwhile pursuit.

The remainder of this paper is outlined as follows.
\S \ref{sec:Observations} lists the new and updated data used in this re-analysis of OB110462. 
\S \ref{sec:Updated astrometric reductions} describes the new and updated capabilities of the \onepass\, software and their effect on the measurement of source positions and magnitudes.
\S \ref{sec:Astrometric alignment using re-analyzed data} describes how these updated measurements are used to derive an updated astrometric time series, and \S \ref{sec:Modeling using updated data} describes how both the updated photometry and astrometry are fit with a microlensing model.
\S \ref{sec:Results} presents the lens' properties, compares them to previous studies of OB110462, and shows that the choice of software significantly affects the astrometry and in turn the lens mass.
\S \ref{sec:Discussion} discusses OB110462 is the context of the known Galactic BH population and considers future searches for BHs.
\S \ref{sec:Conclusion} provides a summary and conclusions.

\section{Observations \label{sec:Observations}}

\subsection{HST \label{sec:HST}}

11 epochs of HST observations of OB110462 were obtained between 2011 and 2022, and are presented in Table \ref{tab:hst_obs}.
This includes all the data analyzed in \cite{Lam:2022a,Lam:2022b} (see Table 2 in \cite{Lam:2022b}), with several additions.
First, all exposures taken on 2011-08-08 were included in this analysis (compared to \cite{Lam:2022a,Lam:2022b}, who excluded 3 frames that had different exposure times than the main dataset).
The exposures from 2013-05-13 was also included (compared to \cite{Lam:2022a,Lam:2022b}, who excluded them because there were no other Spring epochs to calibrate reference frame alignment issues due to parallax; however there is now a second Spring epoch that can be used to perform this calibration).
Finally, the new HST observations taken in 2022 have been included.
Note that although the 2022-09-13 GO data was taken with the UVIS2-2K2C-SUB subarray like the previous datasets, the 2022-05-29 SNAP data was taken with the UVIS2-C1K1C-SUB subarray, which is a smaller subarray (1k$\times$1k, vs. 2k$\times$2k).

\begin{deluxetable}{lcccc}
\tablecaption{HST data analyzed
\label{tab:hst_obs}}
\tablehead{
    \colhead{Epoch} & 
    \colhead{PA} & 
    \colhead{Filter} &
    \colhead{$T_{exp}$} & 
    \colhead{$N_{im}$} \\
    \colhead{(UT)} & 
    \colhead{(deg)} & 
    &
    \colhead{(sec)} & 
    }
\startdata
2011-08-08 & 270.0 & F606W & 75.0 & 3 \\ 
\textbf{2011-08-08} & \textbf{270.0} & \textbf{F606W} & \textbf{60.0} & \textbf{1} \\ 
2011-08-08 & 270.0 & F814W & 75.0 & 3 \\ 
\textbf{2011-08-08} & \textbf{270.0} & \textbf{F814W} & \textbf{60.0} & \textbf{1} \\ 
\textbf{2011-08-08} & \textbf{270.0} & \textbf{F814W} & \textbf{120.0} & \textbf{1} \\ 
\hline 
2011-10-31 & 276.1 & F606W & 280.0 & 3 \\ 
2011-10-31 & 276.1 & F814W & 200.0 & 4 \\ 
\hline 
2012-09-09 & 269.5 & F606W & 290.0 & 3 \\ 
2012-09-09 & 269.5 & F814W & 190.0 & 4 \\ 
\hline 
2012-09-25 & 271.3 & F606W & 280.0 & 3 \\ 
2012-09-25 & 271.3 & F814W & 200.0 & 4 \\ 
\hline 
\textbf{2013-05-13} & \textbf{99.9} & \textbf{F606W} & \textbf{280.0} & \textbf{3} \\ 
\textbf{2013-05-13} & \textbf{99.9} & \textbf{F814W} & \textbf{200.0} & \textbf{4} \\ 
\hline 
2013-10-22 & 274.6 & F606W & 285.0 & 3 \\ 
2013-10-22 & 274.6 & F814W & 285.0 & 4 \\ 
\hline 
2014-10-26 & 275.2 & F606W & 265.0 & 3 \\ 
2014-10-26 & 275.2 & F814W & 265.0 & 4 \\ 
\hline 
2017-08-29 & 268.3 & F606W & 250.0 & 3 \\ 
2017-08-29 & 268.3 & F814W & 250.0 & 4 \\ 
\hline 
2021-10-01 & 272.0 & F606W & 407.0 & 5 \\ 
2021-10-01 & 272.0 & F814W & 307.0 & 6 \\ 
\hline 
\textbf{2022-05-29} & \textbf{107.9} & \textbf{F814W} & \textbf{300.0} & \textbf{2} \\ 
\hline 
\textbf{2022-09-13} & \textbf{269.9} & \textbf{F606W} & \textbf{407.0} & \textbf{5} \\ 
\textbf{2022-09-13} & \textbf{269.9} & \textbf{F814W} & \textbf{307.0} & \textbf{6}

\enddata
\tablecomments{HST data.
For each epoch, the position angle (PA), HST WFC3-UVIS filter, exposure time $T_{exp}$, and number of images $N_{im}$ are listed.
\emph{Bold text} indicates data used in this re-analysis that was not used in \cite{Lam:2022a,Lam:2022b}.
}
\end{deluxetable}

\subsection{OGLE \label{sec:OGLE}}

\begin{figure*}[t]
    \centering
    \includegraphics[width=\linewidth]{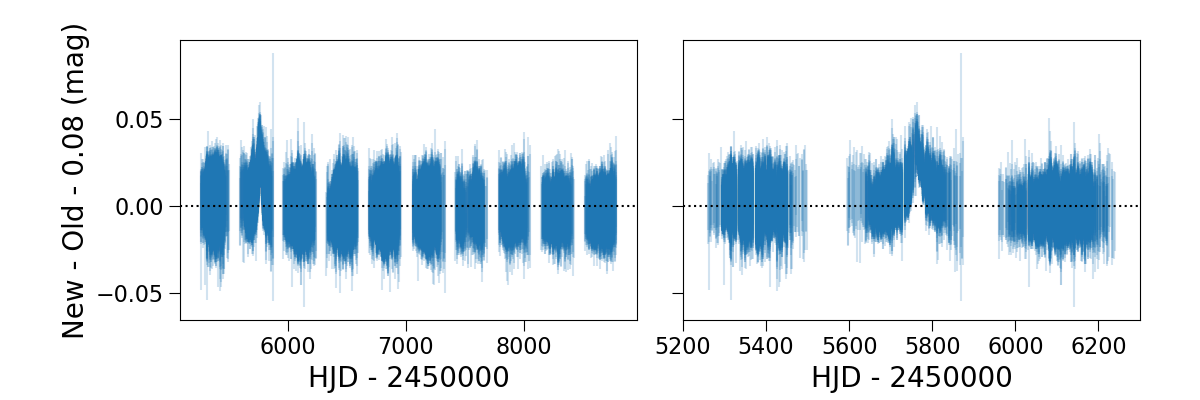}
    \caption{\label{fig:lightcurve_difference}
    Difference between the OGLE photometry used in \cite{Mroz:2022} as to that used in \cite{Lam:2022a, Lam:2022b, Sahu:2022}.
    \emph{Left:} full lightcurve.
    \emph{Right:} lightcurve zoomed into the first three years (photometric peak year $\pm$ 1 year).
    }
\end{figure*}

As mentioned in \S \ref{sec:Introduction}, the ground-based OGLE photometry was re-reduced by \cite{Mroz:2022}.
Figure \ref{fig:lightcurve_difference} shows the difference between the old and new lightcurve.
In addition, \cite{Mroz:2022} found that data from the first half of 2010 was affected by systematics due to commissioning of a new camera, and removed this data from their analysis.
They also only modeled data through 2016 as they found a potential systematic in old OGLE reductions around HJD = 2458000 (September 2017).
For consistency, we also model the same subset of re-reduced OGLE data, spanning HJD = 2455376 to 2457700 (roughly July 2010 - November 2016).

\section{Updated astrometric reductions and analysis with \onepass \label{sec:Updated astrometric reductions}}

The software package \onepass\, extracts precise astrometry from HST WFC3-UVIS imaging.
It is described in \cite{Anderson:2006}, and although it has been updated and used in many publications over the years, never formally released as Space Telescope Science Institue-supported software.
\cite{Lam:2022a, Lam:2022b} and \cite{Sahu:2022} used one of these unofficial \onepass\, releases to perform their astrometric analyses.

In July 2022 \onepass\, was officially released \citep{Anderson:2022}.
Most notably, this release included a tabular correction for charge transfer efficiency \citep[CTE,][]{Anderson:2021_CTEtab} that could be used instead of a pixel-based CTE correction \citep{Anderson:2021_CTEv2.0},
and new functionality to perform artificial star injection and recovery simulations.

We note that the official release of \onepass\, also comes with a new routine called \collate\, to collate the starlists of the individual frames together into a final starlist for that epoch. 
However, because this is a limited-use early version of \collate, we find the flexibility of the existing software routines \texttt{xym2mat} and \texttt{xym2bar} \citep{Anderson:2006} to be superior.
Hence, we do not use \collate\, in our analysis and do not discuss it further here.

\subsection{CTE correction \label{sec:new CTE correction}}

\cite{Lam:2022a, Lam:2022b} and \cite{Sahu:2022} used the CTE-corrected flat-fielded HST images (i.e. \texttt{flc}) for their analyses. 
The \texttt{flc} images were produced using Version 2.0 of a pixel-based CTE correction algorithm \citep{Anderson:2021_CTEv2.0}.
However, the pixel-based correction usually under-corrects the CTE effect on photometry \citep{Kuhn:2021}. 

The updated version of \onepass\, includes a tabular correction that empirically corrects for CTE based on the brightness of the source and the sky background, which improves the extraction of photometry and astrometry \citep{Anderson:2021_CTEtab}.

Both CTE correction methods alter the extracted source positions in the detector $y$ direction, which is the parallel readout direction.
At present, CTE in the detector $x$ direction, which is the serial readout direction, is not corrected.
Although there is CTE in the serial readout direction, it is negligibly small compared to CTE in the parallel readout direction \citep{Anderson:2014}. 

When performing the data reduction in this work, instead of reducing the CTE pixel-corrected \texttt{flc} files, we instead reduce the flat-fielded data files (i.e. \texttt{flt}) with \onepass\, using the tabular CTE correction.
Measured positions between these two methods can differ by a tenth of a pixel and measured brightness can differ by a tenth of a magnitude (Figure \ref{fig:cte} in Appendix \ref{app:CTE}).
We then proceed with the data reduction and intra-epoch alignment process described in \S 4.1 of \cite{Lam:2022b} using the tabular-CTE corrected starlists.

\subsection{Artificial star injection and recovery tests}

OB110462 is located $\sim$10 pixels ($\sim$0.4 arcsec) away from an unrelated neighbor star that is 3 magnitudes brighter.
This neighbor star biases the measurement of the flux and position of OB110462.
\cite{Lam:2022a, Lam:2022b} and \cite{Sahu:2022} took different approaches to calculate this bias.

\cite{Sahu:2022} used 18 nearby isolated stars with color and magnitude comparable to the neighbor to construct an ``extended model" point spread function (PSF).
This extended model PSF was then subtracted from each exposure to obtain an unbiased position and magnitude of OB110462.
Using this method, \cite{Sahu:2022} found that the typical positional bias for OB110462 was about 1.2 mas.

\cite{Lam:2022a, Lam:2022b} performed injection and recovery tests to measure the bias.
They injected sources around an isolated star of similar brightness to the neighbor, at the same azimuth, separation, and magnitude difference as the neighbor-OB110462 pair.
Using this method, \cite{Lam:2022b} found a smaller positional bias of around 0.3 and 0.5 mas for OB110462, in the F606W and F814W filters, respectively.

In \S \ref{sec:1pass vs ks2} we describe the new \onepass\, software used to perform source extraction for artificial stars and in \S \ref{sec:Updated methodology}, we present an updated and more extensive star-planting analysis using the new \onepass.

\subsubsection{\onepass\, vs. \kstwo\, software \label{sec:1pass vs ks2}}

The version of \onepass\, used by \cite{Lam:2022a,Lam:2022b} did not have a method to generate artificial stars in the images.
Thus, \cite{Lam:2022a,Lam:2022b} used a different software package, called \kstwo, to perform the injection and recovery tests.
\kstwo\, has not been formally released, but it is described in several papers, e.g. \cite{Anderson:2008, Bellini:2018, Sabbi:2016}.

Although both \onepass\, and \kstwo\, are used to extract precise astrometric measurements from HST imaging, they work in slightly different manners.
\kstwo\, was specifically designed to find fainter sources than \onepass.
There are also certain implementation differences across the two software packages.
Of relevance to the astrometry are the geometric distortion solutions used, and the specifics of the PSF fitting.
With regard to the distortion solution, \kstwo\, has an internal geometric distortion solution that is slightly different from the standard geometric distortion correction (STDGDC) files used by \onepass.
With regard to the PSFs, the specific manner of fitting slightly differ, e.g. the particulars of how outlier rejection is implemented.

In \cite{Lam:2022a,Lam:2022b}, \onepass\, was used instead of \kstwo\, for data reduction because recovery depth was not an issue 
and \onepass\, had been more robustly used and tested on HST WFC3-UVIS data.
To use \onepass\, to obtain the positions, and then use \kstwo\, to calculate the bias in the position and flux of OB110462 is formally inconsistent, since the two methods of source extraction in these software differ.
As a validation test, \onepass\, and \kstwo\, were used to extract astrometry from the same epoch of HST observations and then the differences between the resultant starlists were compared.
No clear trends were found to explain the differences.
Since the bias correction is a relative measurement and self-consistent across one software package, it was deemed a reasonable approach to calculate the bias correction with \kstwo\, and then apply it to measurements made with \onepass.
However, now that the new version of \onepass\, has the ability to simulate artificial stars, the injection and recovery analysis can be done in a fully self-consistent manner and this assumption can be checked.

\subsubsection{Updated methodology \label{sec:Updated methodology}}

\begin{figure}[t!]
    \centering
    \includegraphics[width=1.0\linewidth]{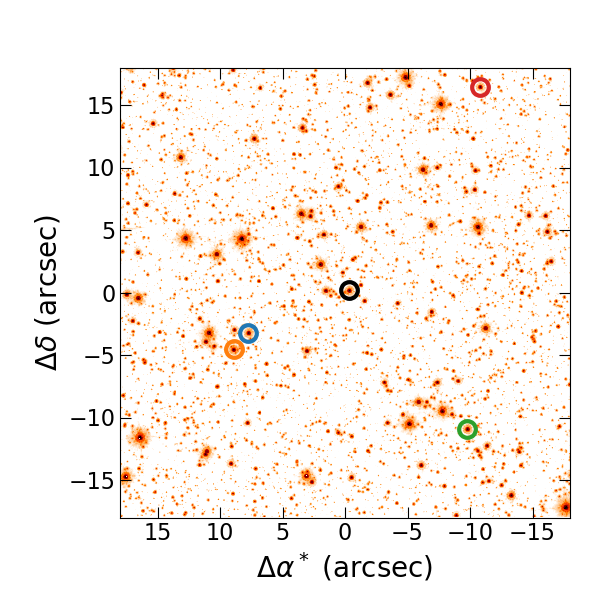}
    \caption{\label{fig:artstar_all}
    Spatial distribution of neighbor-like stars.
    The image shows a $36 \times 36$ arcsec area centered on OB110462.
    The black circle in the center shows the location of OB110462's bright neighbor.
    The four other colored circles show the location of the neighbor-like stars.
    }
\end{figure}

\begin{figure}[t!]
    \centering
    \includegraphics[width=1.0\linewidth]{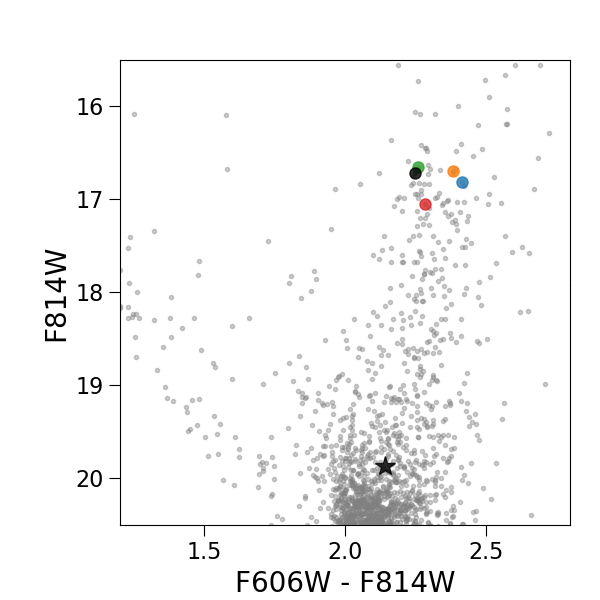}
    \caption{\label{fig:artstar_cmd}
    Location of neighbor-like stars on an HST CMD.
    The black circle marks OB110462's bright neighbor.
    The four other colored circles mark the neighbor-like stars (colors correspond to those in Figure \ref{fig:artstar_all}).
    The black star marks OB110462 at baseline (i.e. unmagnified).
    }
\end{figure}

\begin{figure*}[t!]
    \centering
    \includegraphics[width=1.0\linewidth]{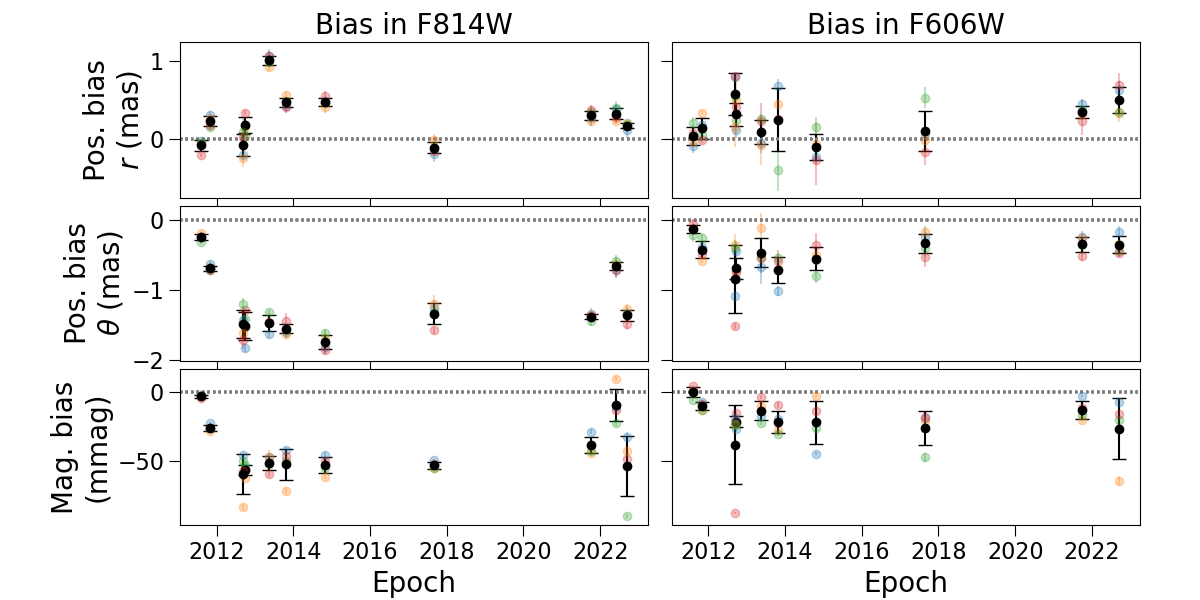}
    \caption{\label{fig:inject_recover_bias}
    Measurement bias of a source in F814W (\emph{left}) and F606W (\emph{right}) due to proximity to a bright star. 
    The bias correction is defined as the recovered minus injected value.
    The amount of positional bias is shown decomposed into radial $r$ (\emph{top row}) and azimuthal $\theta$ (\emph{middle row}) components .
    The radial direction is defined by the OB110462-neighbor star separation vector, and the azimuthal direction is measured counterclockwise from the separation vector.
    The amount of magnitude bias is shown in the \emph{bottom row}.
    The colored points show the mean and standard deviation of the measurement bias for the four different neighbor-like stars; the colors correspond to those in Figure \ref{fig:artstar_all} and Figure \ref{fig:artstar_cmd}.
    The black points are the mean and standard deviation of the mean measurement bias of the four neighbor-like stars.
    The positional and magnitude bias in F606W is smaller than that in F814W by about a factor of two; this is not surprising since at shorter wavelength the angular resolution is higher.
    }
\end{figure*}

\begin{deluxetable}{lcccc}
\tablecaption{Bias correction derived from injection and recovery
\label{tab:injection recovery bias_eom}}
\tablehead{
    \colhead{Epoch} & 
    \colhead{$\Delta$RA (mas)} & 
    \colhead{$\Delta$Dec (mas)} &
    \colhead{$\Delta$Total (mas)} &
    \colhead{$\Delta$Mag (mag)}}
\startdata
F606W & & & & \\
\hline
2011-08-08 & -0.14 $\pm$ 0.08 & 0.00 $\pm$ 0.10 & 0.14 & 0.000 $\pm$ 0.004 \\ 
2011-10-31 & -0.46 $\pm$ 0.14 & 0.00 $\pm$ 0.11 & 0.46 & -0.010 $\pm$ 0.003 \\ 
2012-09-09 & -0.98 $\pm$ 0.53 & -0.27 $\pm$ 0.15 & 1.02 & -0.039 $\pm$ 0.029 \\ 
2012-09-25 & -0.76 $\pm$ 0.17 & -0.08 $\pm$ 0.12 & 0.76 & -0.022 $\pm$ 0.004 \\ 
2013-05-13 & -0.48 $\pm$ 0.22 & 0.06 $\pm$ 0.13 & 0.48 & -0.013 $\pm$ 0.007 \\ 
2013-10-22 & -0.76 $\pm$ 0.29 & -0.01 $\pm$ 0.33 & 0.76 & -0.022 $\pm$ 0.008 \\ 
2014-10-26 & -0.50 $\pm$ 0.20 & 0.27 $\pm$ 0.12 & 0.57 & -0.022 $\pm$ 0.016 \\ 
2017-08-29 & -0.36 $\pm$ 0.15 & 0.00 $\pm$ 0.25 & 0.36 & -0.026 $\pm$ 0.012 \\ 
2021-10-01 & -0.44 $\pm$ 0.08 & -0.22 $\pm$ 0.10 & 0.49 & -0.013 $\pm$ 0.007 \\ 
2022-09-13 & -0.50 $\pm$ 0.12 & -0.36 $\pm$ 0.16 & 0.62 & -0.027 $\pm$ 0.022 \\ 

\\\hline
F814W & & & &\\
\hline
2011-08-08 & -0.21 $\pm$ 0.05 & 0.15 $\pm$ 0.07 & 0.26 & -0.003 $\pm$ 0.001 \\ 
2011-10-31 & -0.73 $\pm$ 0.02 & 0.00 $\pm$ 0.07 & 0.73 & -0.026 $\pm$ 0.002 \\ 
2012-09-09 & -1.39 $\pm$ 0.18 & 0.55 $\pm$ 0.17 & 1.49 & -0.060 $\pm$ 0.015 \\ 
2012-09-25 & -1.50 $\pm$ 0.18 & 0.31 $\pm$ 0.13 & 1.53 & -0.057 $\pm$ 0.003 \\ 
2013-05-13 & -1.72 $\pm$ 0.12 & -0.47 $\pm$ 0.05 & 1.78 & -0.052 $\pm$ 0.005 \\ 
2013-10-22 & -1.63 $\pm$ 0.08 & 0.05 $\pm$ 0.05 & 1.63 & -0.053 $\pm$ 0.012 \\ 
2014-10-26 & -1.81 $\pm$ 0.10 & 0.11 $\pm$ 0.03 & 1.81 & -0.053 $\pm$ 0.006 \\ 
2017-08-29 & -1.23 $\pm$ 0.13 & 0.53 $\pm$ 0.10 & 1.34 & -0.053 $\pm$ 0.002 \\ 
2021-10-01 & -1.41 $\pm$ 0.05 & 0.15 $\pm$ 0.05 & 1.42 & -0.039 $\pm$ 0.006 \\ 
2022-05-29 & -0.74 $\pm$ 0.06 & -0.10 $\pm$ 0.07 & 0.74 & -0.009 $\pm$ 0.012 \\ 
2022-09-13 & -1.35 $\pm$ 0.07 & 0.27 $\pm$ 0.04 & 1.37 & -0.054 $\pm$ 0.022 \\ 

\enddata
\tablecomments{
Bias correction derived from injection/recovery around a star of comparable brightness at the same separation, azimuth, and magnitude difference as OB110462 to its bright neighbor.
The bias correction is defined as the recovered minus the true injected value.
}
\end{deluxetable}

Here, we update the analysis performed in \cite{Lam:2022b}.
We briefly summarize the methodology here and only highlight new changes; see Appendix B of \cite{Lam:2022b} for full details.

We calculate the bias in the position and flux of OB110462 by injecting artificial stars using \onepass\, at the same azimuth and separation as OB110462 and its bright neighbor star, around ``neighbor-like" stars.
We then determine whether these injected artificial stars are recovered, and if they are, how different the recovered and injected positions and fluxes are.

The criteria for selecting nearby isolated neighbor-like stars are:
\begin{itemize}
    \item similar brightness to neighbor (within $\pm$ 0.5 mag in both F814W and F606),
    \item similar color to neighbor (within $\pm$ 0.25 mag in F606W $-$ F814W),
    \item nearby to neighbor (within $\pm$20 arcsec = 500 pix)
    \item isolated from other stars (at least 0.4 arcsec = 10 pix away from any other source detected by \onepass\, in F814W).
\end{itemize}
There are four stars that fit all these criteria. 
Their positions in relation to OB110462 are shown in Figure \ref{fig:artstar_all} and on a color-magnitude diagram (CMD) in Figure \ref{fig:artstar_cmd}.

45 artificial stars are injected in a $0.2 \times 0.2$ pixel ($ = 8 \times 8$ mas) area adjacent to each of the four neighbor-like stars, with a magnitude so that the artificial star has the same contrast with the neighbor-like star as OB110462 to the neighbor.
The results of the injection and recovery are shown in Figures \ref{fig:inject_recover_bias} and \ref{fig:inject_recover_bias_comparison} (c.f. Figures 22 and 23 in \cite{Lam:2022b}) and listed in Table \ref{tab:injection recovery bias_eom} (c.f. Table 16 in \cite{Lam:2022b}).

\section{Updated cross-epoch alignment \label{sec:Astrometric alignment using re-analyzed data}}

\begin{figure}[h!]
    \centering
    \includegraphics[width=\linewidth]{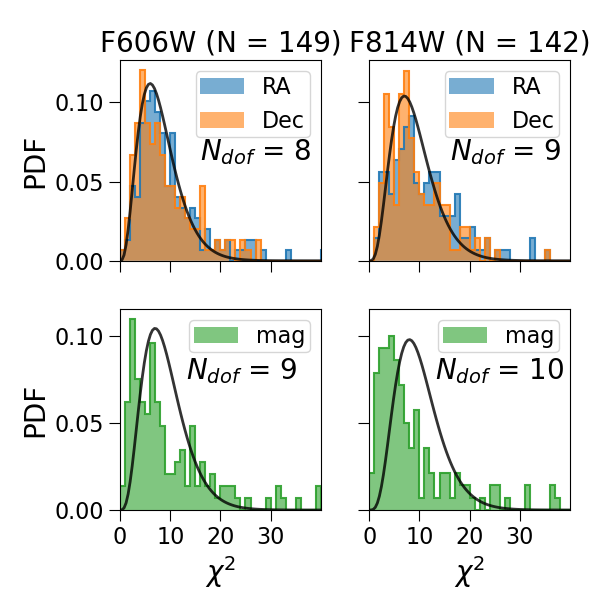}
    \caption{\label{fig:chi2_xym}
    \emph{Top:} Histogram of $\chi^2$ residual values to linear proper motion fits (no parallax) of the reference stars. 
    \emph{Bottom:} Histogram of $\chi^2$ residual values to constant magnitude vs. time fits of the reference stars. 
    The \emph{left} column shows the reference stars for F606W; the \emph{right} column shows the reference stars for F814W.
    $N$ denotes the number of reference stars.}
\end{figure}

\begin{figure*}[h!]
    \centering
    \includegraphics[angle=-90,width=0.8\linewidth]{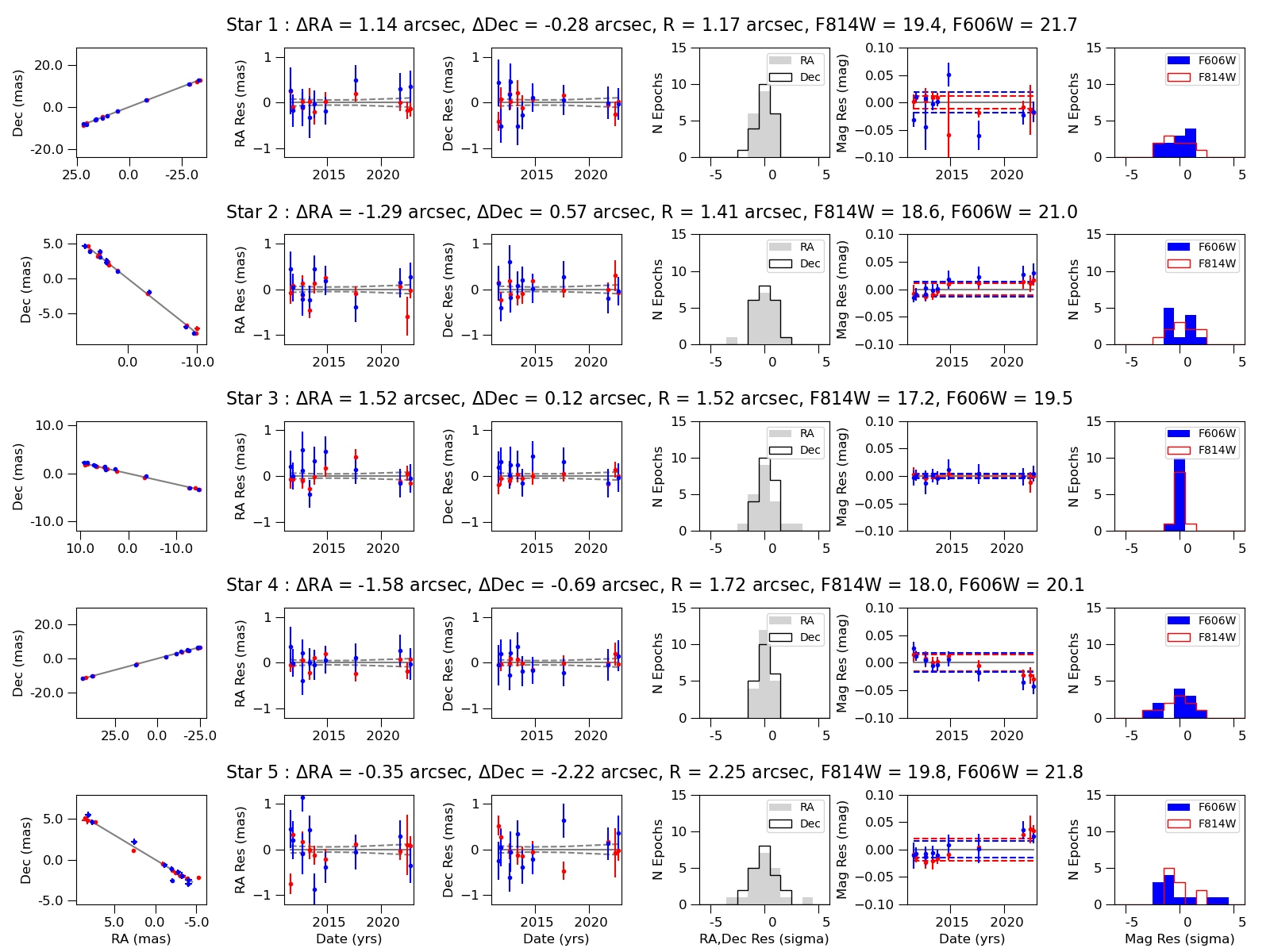}
    \caption{\label{fig:by_eye_ref_stars1}
    5 reference stars nearest to OB110462.
    The title indicates the separation of the reference star from OB110462 in RA, Dec, and total distance, and the magnitude of the star in F814W and F606W.
    In the individual panels, data in the F606W (F814W) filters are shown in \emph{blue} (\emph{red}).
    \emph{1st column:} Trajectory on sky. 
    Note the scales across different rows vary.
    The positions are relative to the $\Delta$RA and $\Delta$Dec offsets, which are relative to OB110462.
    The best-fit linear trajectory is shown in \emph{gray}.
    Note these velocities are in a reference frame where the mean velocity of reference stars is 0, and not the Gaia reference frame.
    \emph{2nd column:} Residuals to the best-fit linear trajectory in RA. 
    The dashed lines are the $1\sigma$ uncertainties to the best-fit line.
    \emph{3rd column:} Same as 2nd column, but for Dec instead of RA.
    \emph{4th column:} Histogram of position residuals to the best-fit linear trajectory, in units of sigma.
    The solid gray is for RA; the outlined black is for Dec.
    \emph{5th column:} Residuals to the best-fit constant magnitude.
    F814W and F606W are fit to the respective magnitudes in the title; the dashed lines show the $1\sigma$ uncertainties to the best-fit magnitude.
    \emph{6th column:} Residuals to the best-fit constant magnitude, in units of sigma.
    The solid blue is for F606W; the outlined red is for F814W.
    }
\end{figure*}

\begin{figure*}[h!]
    \centering
    \includegraphics[angle=-90,width=0.8\linewidth]{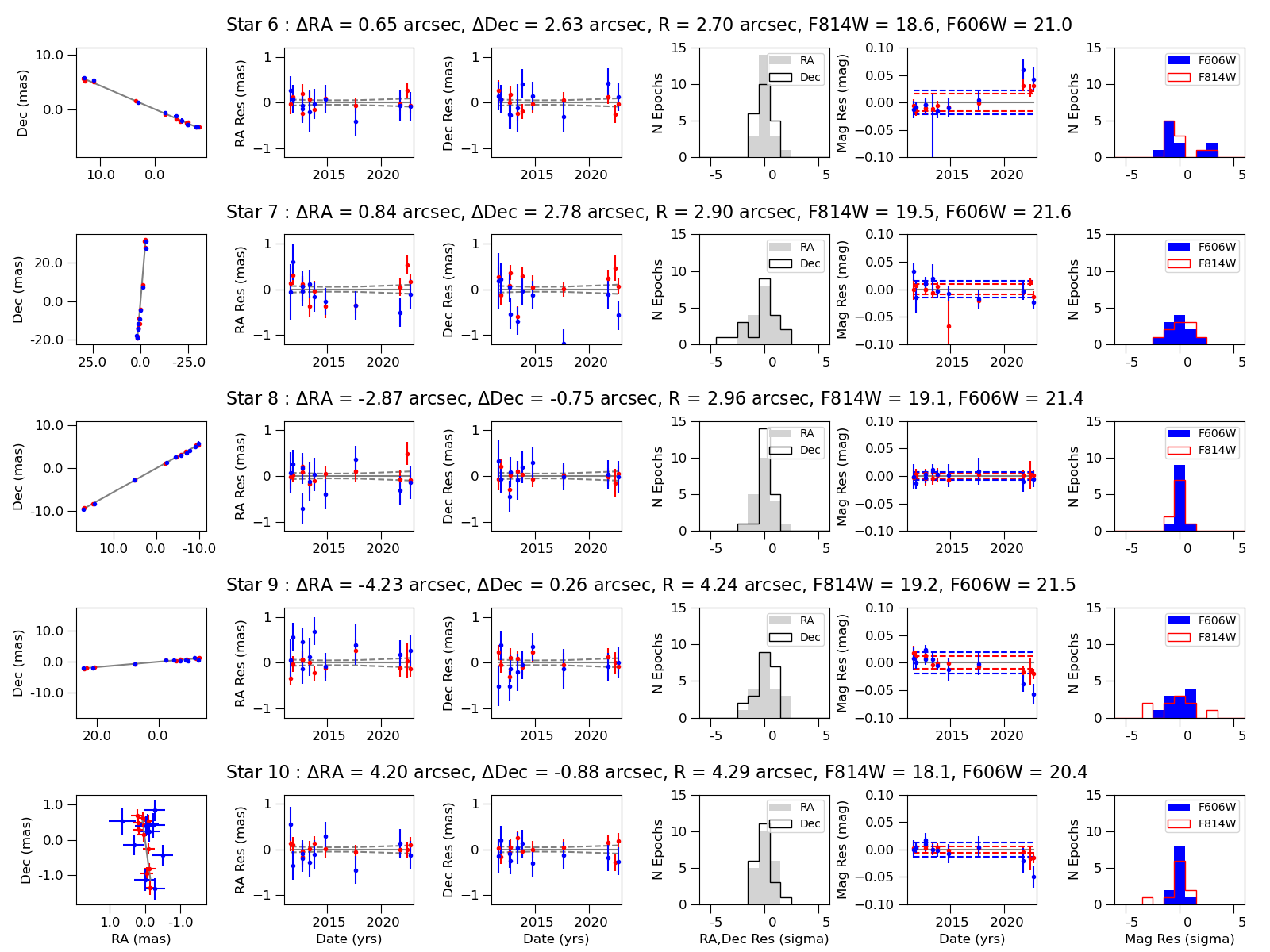}
    \caption{\label{fig:by_eye_ref_stars2}
    Same as Figure \ref{fig:by_eye_ref_stars1}, but for the next 5 reference stars nearest to OB110462.
    }
\end{figure*}

We take the starlists for epoch obtained in \S \ref{sec:new CTE correction} and align them onto a common reference frame, as described in \S 4.2 - 4.2.3 of \cite{Lam:2022b}.
The photometry is calibrated as described in \S 4.3 of \citet{Lam:2022b}; in short, to obtain precise relative photometry, we calculate and apply a small magnitude offset that assumes the reference stars are constant brightness.
One minor change in this work as compared to \cite{Lam:2022b} is the value of the additive error added in quadrature to the positional and magnitude uncertainties.
Appendix A of \cite{Lam:2022b} describes an empirical methodology to calculate the rescaling factor, which can vary epoch to epoch.
We instead simply use a constant additive error across all epochs; the value of the additive error is chosen to make the alignment residuals follow the expected $\chi^2$ distribution.
We find an additive error of 0.25 mas and 12 milli-mag added to the F606W positions and magnitudes, and 0.10 mas and 8 milli-mag in F814W, produce acceptable $\chi^2$ distributions by eye (Figure \ref{fig:chi2_xym}, c.f. Figure 9 of \cite{Lam:2022b}.).
Figures \ref{fig:by_eye_ref_stars1} and \ref{fig:by_eye_ref_stars2} show the 10 astrometric reference stars closest to OB110462.

\begin{figure*}
    \centering
    \includegraphics[width=1.0\linewidth]{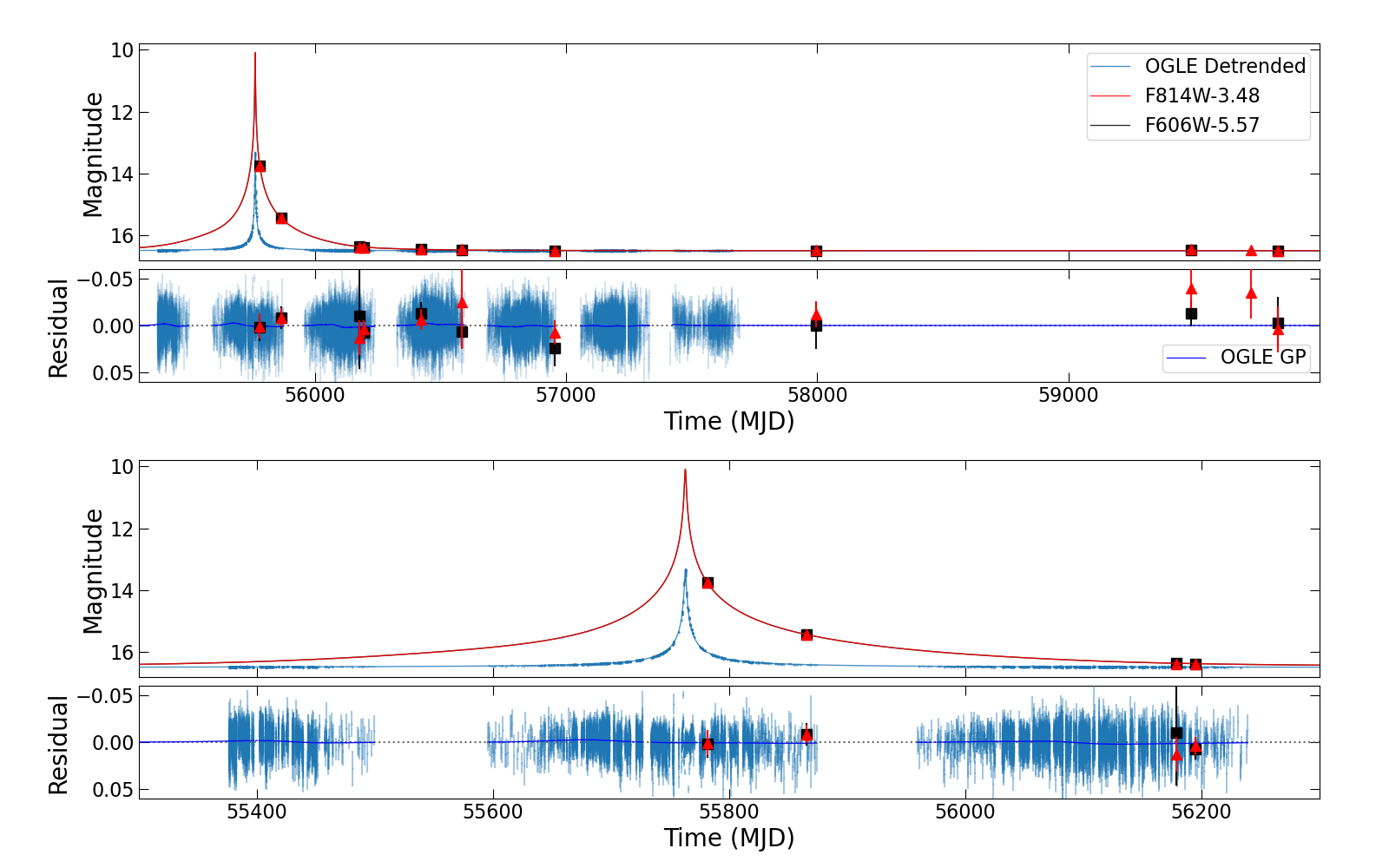}
    \caption{\label{fig:OB110462_DW_lightcurve}
    \emph{Top panel}: HST F814W \emph{red triangles}, HST F606W \emph{black squares}, and detrended OGLE lighturve \emph{blue} lightcurves, with the corresponding  maximum likelihood model (MLE, described in \S \ref{sec:Modeling using updated data}) plotted over the data.
    \emph{Second from top panel}: Residuals to the MLE model.
    The Gaussian Process (GP) model is plotted on top of the OGLE residual.
    \emph{Second from bottom panel}: Same as top panel, but zoomed into the three most magnified years (2010-2012).
    \emph{Bottom panel}: Same as second from top panel, but zoomed into the three most magnified years  (2010-2012).}
\end{figure*}

\begin{figure*}
    \centering
    \includegraphics[width=1.0\linewidth]{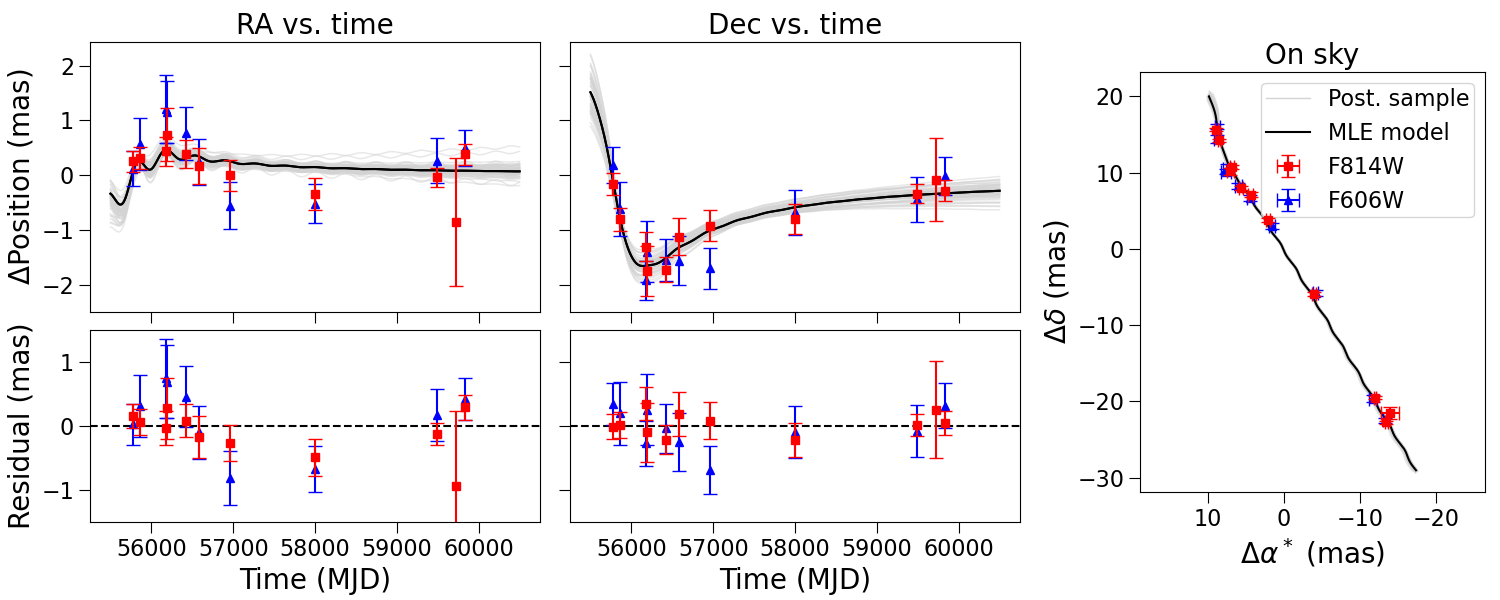}
    \caption{\label{fig:DW_astrometry}
    OB110462 astrometry.
    \emph{Left column, top to bottom}: RA vs. time with maximum likelihood (MLE) unlensed source motion model subtracted; residuals to the MLE model for RA vs. time fit.
    HST F814W astrometry data is shown in red; HST F606W astrometry data is shown in blue.
    The MLE model is shown in black.
    Fifty random draws from the posterior distribution are shown in light gray.
    \emph{Middle column, top to bottom}: Same as left column, except Dec instead of RA.
    \emph{Right panel}: astrometry as seen on-sky, in the barycentric frame.
    }
\end{figure*}

The final result of the cross-epoch alignment is the HST photometric and astrometric time series (Figures \ref{fig:OB110462_DW_lightcurve} and \ref{fig:DW_astrometry}).
The bias correction measured in \S \ref{sec:Updated methodology} is then applied to OB110462 to obtain its true positions and magnitudes (Table \ref{tab:hst_data}).
Note that in Table \ref{tab:hst_data} the reported uncertainties do not include the uncertainties in the transformation from a relative astrometric reference frame to the absolute Gaia reference frame, which are 0.13 mas/yr and 0.11 mas/yr in RA and Dec, respectively.

\begin{deluxetable}{lccc}
\tablecaption{HST calibrated data
\label{tab:hst_data}}
\tablehead{
    \colhead{Epoch} & 
    \colhead{RA (mas)} &
    \colhead{Dec (mas)} & 
    \colhead{Mag (Vega)}}
\startdata
F814W & & & \\ 
\hline 
2022-09-13 & -13.43 $\pm$ 0.19 & -22.75 $\pm$ 0.19 & 19.965 $\pm$ 0.024 \\ 
2022-05-29 & -13.91 $\pm$ 1.17 & -21.55 $\pm$ 0.76 & 19.926 $\pm$ 0.027 \\ 
2021-10-01 & -11.93 $\pm$ 0.17 & -19.53 $\pm$ 0.17 & 19.921 $\pm$ 0.021 \\ 
2017-08-29 & -3.97 $\pm$ 0.29 & -5.88 $\pm$ 0.27 & 19.949 $\pm$ 0.014 \\ 
2014-10-26 & 2.11 $\pm$ 0.29 & 3.81 $\pm$ 0.28 & 19.964 $\pm$ 0.013 \\ 
2013-10-22 & 4.32 $\pm$ 0.33 & 7.10 $\pm$ 0.34 & 19.918 $\pm$ 0.050 \\ 
2013-05-13 & 5.62 $\pm$ 0.26 & 8.01 $\pm$ 0.23 & 19.916 $\pm$ 0.011 \\ 
2012-09-25 & 7.04 $\pm$ 0.48 & 10.17 $\pm$ 0.46 & 19.849 $\pm$ 0.010 \\ 
2012-09-09 & 6.83 $\pm$ 0.27 & 10.76 $\pm$ 0.26 & 19.846 $\pm$ 0.018 \\ 
2011-10-31 & 8.47 $\pm$ 0.20 & 14.23 $\pm$ 0.21 & 18.897 $\pm$ 0.009 \\ 
2011-08-08 & 8.87 $\pm$ 0.19 & 15.65 $\pm$ 0.20 & 17.230 $\pm$ 0.014 \\ 
\hline 
F606W & & & \\ 
\hline 
2022-09-13 & -13.31 $\pm$ 0.33 & -22.48 $\pm$ 0.35 & 22.054 $\pm$ 0.028 \\ 
2021-10-01 & -11.63 $\pm$ 0.41 & -19.63 $\pm$ 0.41 & 22.043 $\pm$ 0.014 \\ 
2017-08-29 & -4.15 $\pm$ 0.36 & -5.76 $\pm$ 0.41 & 22.056 $\pm$ 0.025 \\ 
2014-10-26 & 1.56 $\pm$ 0.42 & 3.03 $\pm$ 0.38 & 22.076 $\pm$ 0.019 \\ 
2013-10-22 & 4.39 $\pm$ 0.42 & 6.66 $\pm$ 0.45 & 22.046 $\pm$ 0.018 \\ 
2013-05-13 & 5.99 $\pm$ 0.48 & 8.18 $\pm$ 0.39 & 22.005 $\pm$ 0.012 \\ 
2012-09-25 & 7.45 $\pm$ 0.57 & 10.52 $\pm$ 0.56 & 21.948 $\pm$ 0.011 \\ 
2012-09-09 & 7.59 $\pm$ 0.62 & 10.15 $\pm$ 0.35 & 21.918 $\pm$ 0.057 \\ 
2011-10-31 & 8.72 $\pm$ 0.48 & 14.41 $\pm$ 0.49 & 20.992 $\pm$ 0.012 \\ 
2011-08-08 & 8.74 $\pm$ 0.32 & 16.00 $\pm$ 0.33 & 19.326 $\pm$ 0.015
\enddata
\tablecomments{Relative positions and magnitudes of OB110462.}
\end{deluxetable}

\section{Modeling using updated data \label{sec:Modeling using updated data}}

\begin{deluxetable}{lccc}
\tablecaption{Posterior distributions of fit parameters
\label{tab:DW fit}}
\tablehead{
    \colhead{Parameter} &
    \colhead{Med$^{+1\sigma}_{-1\sigma}$} & 
    \colhead{MAP} &
    \colhead{MLE}}
\startdata
$t_0$ (MJD) & $55764.47_{{-0.93}}^{{+0.85}}$&55764.69&55763.84\\ 
$u_0$ & $-0.05_{{-0.007}}^{{+0.007}}$&-0.05&-0.05\\ 
$t_E$ (days) & $275.98_{{-6.01}}^{{+5.23}}$&267.26&287.48\\ 
$\log_{10}(\theta_E$/mas) & $0.68_{{-0.06}}^{{+0.05}}$&0.68&0.69\\ 
$\pi_S$ (mas) & $0.11_{{-0.02}}^{{+0.02}}$&0.09&0.12\\ 
$\pi_{E,E}$ & $0.03_{{-0.005}}^{{+0.005}}$&0.03&0.02\\ 
$\pi_{E,N}$ & $-0.09_{{-0.01}}^{{+0.01}}$&-0.10&-0.09\\ 
$x_{S0,E}$ (mas) & $230.31_{{-0.11}}^{{+0.11}}$&230.32&230.35\\ 
$x_{S0,N}$ (mas) & $-214.76_{{-0.16}}^{{+0.16}}$&-214.90&-214.73\\ 
$\mu_{S,E}$ (mas/yr) & $-2.02_{{-0.01}}^{{+0.01}}$&-2.02&-2.02\\ 
$\mu_{S,N}$ (mas/yr) & $-3.45_{{-0.02}}^{{+0.02}}$&-3.45&-3.45\\ 
$b_{SFF,O}$ & $0.05_{{-0.001}}^{{+0.002}}$&0.05&0.05\\ 
$m_{base,O}$ (mag) & $16.48_{{-0.0004}}^{{+0.0004}}$&16.49&16.49\\ 
$b_{SFF,H8}$ & $0.98_{{-0.03}}^{{+0.03}}$&0.93&0.98\\ 
$m_{base,H8}$ (mag) & $19.96_{{-0.005}}^{{+0.005}}$&19.96&19.96\\ 
$b_{SFF,H6}$ & $0.98_{{-0.03}}^{{+0.03}}$&0.95&0.98\\ 
$m_{base,H6}$ (mag) & $22.05_{{-0.006}}^{{+0.006}}$&22.05&22.06\\ 
\hline 
$M_L$ ($M_\odot$) & $6.03_{{-1.04}}^{{+1.19}}$&5.40&6.39\\ 
$\pi_L$ (mas) & $0.58_{{-0.09}}^{{+0.09}}$&0.60&0.58\\ 
$\pi_{rel}$ (mas) & $0.47_{{-0.09}}^{{+0.09}}$&0.52&0.45\\ 
$\mu_{L,E}$ (mas/yr) & $-3.80_{{-0.55}}^{{+0.48}}$&-3.84&-3.50\\ 
$\mu_{L,N}$ (mas/yr) & $2.60_{{-0.80}}^{{+0.83}}$&2.80&2.53\\ 
$\mu_{rel,E}$ (mas/yr) & $1.78_{{-0.48}}^{{+0.56}}$&1.82&1.49\\ 
$\mu_{rel,N}$ (mas/yr) & $-6.05_{{-0.83}}^{{+0.82}}$&-6.25&-5.98\\ 
$\theta_E$ (mas) & $4.79_{{-1.15}}^{{+1.13}}$&4.76&4.85\\ 
$\pi_E$ & $0.10_{{-0.01}}^{{+0.01}}$&0.05&0.10\\ 
$\delta_{c,max}$ (mas) & $1.69_{{-0.41}}^{{+0.40}}$&1.68&1.72\\ 

\enddata
\tablecomments{The columns list the median $\pm1\sigma$ (68\%) credible intervals, maximum a posteriori (MAP) solution, and and maximum likelihood estimator (MLE) solution for the microlensing model parameters.}
\end{deluxetable}

Next, we fit the re-reduced OGLE data and updated HST data following the procedure outlined in \cite{Lam:2022a,Lam:2022b} for their ``default weight" (DW) fit.
See \S 5 of \cite{Lam:2022b} for details on the model parameters and modeling framework.

Simultaneously fitting the photometry and astrometry is very time-intensive, so to speed up the process, we first simultaneously fit the OGLE and HST photometry, accounting for correlated noise while fitting the OGLE photometry with a Gaussian Process.
The joint photometric and astrometric geometric parameters are $t_0$, $u_0$, $t_E$, $\pi_{E,E}$, and $\pi_{E,N}$.
The photometric-only parameters are $b_{SFF,O}$, $m_{base,O}$, $b_{SFF,H8}$, $m_{base,H8}$, $b_{SFF,H6}$, and $m_{base,H6}$.
The astrometric-only parameters are $x_{S0,E}$, $x_{S0,N}$, $\pi_S$, $\log_{10}\theta_E$, $\mu_{S,E}$, and $\mu_{S,N}$.
The Gaussian process parameters are $\log \sigma_{0,O}$, $\rho_O$, $\log \omega^4_{0,O} S_{0,O}$, and $\log \omega_{0,O}$.
See \S 3 of \cite{Lam:2022a}, and \S 5, \S 5.1 and Appendix F of \cite{Lam:2022b} for a full description of all these parameters.

We then take the posterior distributions for $t_0$, $u_0$, $t_E$, $\pi_{E,E}$, and $\pi_{E,N}$ from the photometry fit, and use them as priors when fitting the HST astrometry.
Note the correlations between the 5 parameters are preserved when using them as a prior in the astrometry fit.
Our priors are listed in Table \ref{tab:priors} in Appendix \ref{app:Priors}.

\section{Results \label{sec:Results}}

The best-fit photometry and astrometry models are shown in Figures \ref{fig:OB110462_DW_lightcurve} and \ref{fig:DW_astrometry}, and the posteriors are listed in Table \ref{tab:DW fit}.
We find that that the lens of OB110462 has a mass of $M_L = 6.03^{+1.19}_{-1.04} M_\odot$, is at a distance $D_L = 1.72^{+0.32}_{-0.23}$ kpc, and has transverse velocity $v_{T,L} = 37.61^{+5.12}_{-5.13}$ km/s.

OB110462 cannot be a a high-mass star, and is thus a BH.
In \cite{Lam:2022a,Lam:2022b}, they rule out any possibility of a luminous lens for OB110462, for a lens mass of $M_L = 3.8 M_\odot$ and source flux fractions $b_{SFF} = 0.9$ and 0.94 in F814W and F606W, respectively.
Here, we find $M_L = 6.0 M_\odot$ and source flux fractions $b_{SFF} = 0.98$ in both F814W and F606W.
In both \cite{Lam:2022a,Lam:2022b} and this work, the lens is $D_L = 1.7$ kpc away.
In this work, the mass is significantly higher and the source flux fractions are also slightly higher. 
This means the constraint on a dark lens is much stronger-- given the lens is at some fixed distance, a higher mass star would be much brighter, and a higher source flux fraction would mean there is less excess flux that such a star could hide in.
Thus, without any shadow of a doubt, OB110462 is a BH.

In the next sections we compare these results to those of \cite{Lam:2022a,Lam:2022b}, \cite{Sahu:2022}, and \cite{Mroz:2022}.
Table \ref{tab:summary of analyses} gives a short summary of the differences in the data and models used to analyze OB110462 across these works.
We first compare the inferred lens properties in \S \ref{sec:Comparison of inferred microlensing parameters}, evaluate the goodness-of-fits of the astrometric models in \S \ref{sec:Goodness-of-fits}, then compare the modeled astrometric time series in \S \ref{sec:Bias correction method} to understand the reasons for the differences in the inferred lens properties.

\subsection{Comparison of inferred microlensing parameters \label{sec:Comparison of inferred microlensing parameters}}

\begin{figure*}
    \centering
    \includegraphics[width=\linewidth]{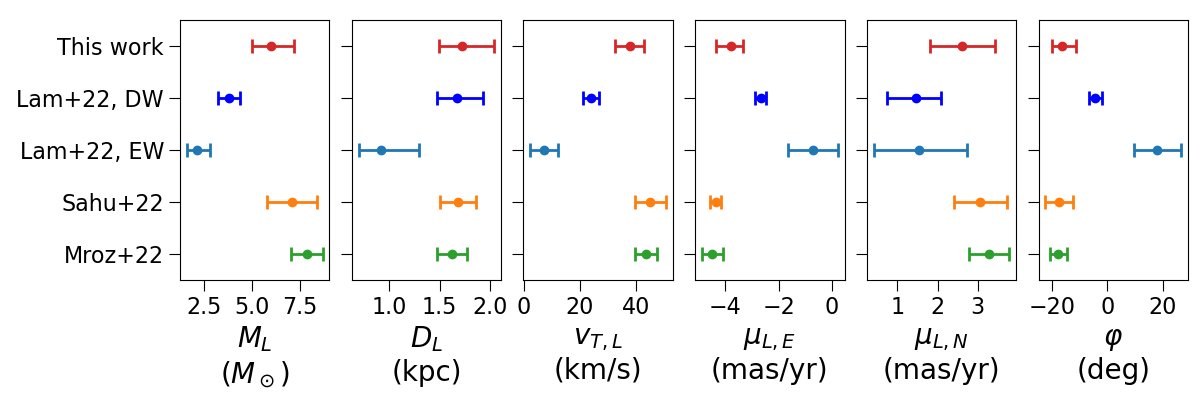} 
    \caption{\label{fig:compare_lens_properties}
    Comparison of lens mass $M_L$, distance $D_L$, transverse velocity $v_{T,L}$, proper motion vector $(\mu_{L,E}$, $\mu_{L,N}$), and lens-source relative proper motion direction $\varphi$ inferred from various studies of OB110462.
    }
\end{figure*}

We compare the inferred lens mass, distance, transverse velocity, and proper motion of OB110462 to \cite{Lam:2022a,Lam:2022b}, \cite{Sahu:2022}, and \cite{Mroz:2022} in Table \ref{tab:comparison} and Figure \ref{fig:compare_lens_properties}.
\cite{Mroz:2022} provides two sets of $D_L$ and $v_{T,L}$ depending on the source distance assumed.
The two solutions are consistent with each other to $1\sigma$; here we only compare to their results using $D_S = 8.8 \pm 1.4$ kpc.
In addition, \cite{Sahu:2022} do not report their uncertainties on the lens transverse velocity; we estimate it to be $\sim 5.5$ km/s based on the reported lens proper motion and distance uncertainties.

The lens mass of OB110462 inferred in this work $M_L = 6.03^{+1.19}_{-1.04} M_\odot$ is consistent with the measurement of \cite{Sahu:2022} $M_L = 7.1 \pm 1.3 M_\odot$ to 1$\sigma$, and consistent with the \cite{Lam:2022a,Lam:2022b} DW model $M_L = 3.79^{+0.62}_{-0.57} M_\odot$ and \cite{Mroz:2022} $M_L = 7.88 \pm 0.82 M_\odot$ measurement to 2$\sigma$ (in different directions).
Our uncertainties are likely larger due to using wider priors for the astrometry; \cite{Mroz:2022} state they use uniform priors in their modeling, but do not state the support.

The lens distance $D_L = 1.72^{+0.32}_{-0.23}$ kpc inferred in this work is consistent with the measurements of \cite{Lam:2022a,Lam:2022b} DW model $D_L = 1.67^{+0.26}_{-0.20}$ kpc, \cite{Sahu:2022} $D_L = 1.58 \pm 0.18$ kpc, and \cite{Mroz:2022} $1.62 \pm 0.15$ kpc to 1$\sigma$.

The lens transverse velocity $v_{T,L} = 37.61^{+5.12}_{-5.13}$ km/s inferred in this work is consistent with \cite{Mroz:2022} $v_{T,L} = 43.4 \pm 3.8$ km/s and \cite{Sahu:2022} $v_{T,L} \sim 45 \pm 5.5$ km/s to 1$\sigma$.
It is consistent with the \cite{Lam:2022a,Lam:2022b} DW model $v_{T,L} = 23.95^{+2.95}_{-2.95}$ km/s to 2$\sigma$.

The lens proper motion $(\mu_{L,E}, \mu_{L,N}) = (-3.80^{+0.48}_{-0.55}, 2.60^{+0.83}_{-0.80})$ mas/yr inferred in this work is consistent with \cite{Mroz:2022} $(\mu_{L,E}, \mu_{L,N}) = (-4.48 \pm 0.39, 3.29 \pm 0.5)$ mas/yr and \cite{Sahu:2022} $(\mu_{L,E}, \mu_{L,N}) = (-4.36 \pm 0.22, 3.06 \pm 0.66)$ mas/yr to 1$\sigma$ in RA and Dec.
The proper motion inferred by the \cite{Lam:2022a,Lam:2022b} DW model $(\mu_{L,E}, \mu_{L,N}) = (-2.64^{+0.18}_{-0.24}, 1.46^{+0.63}_{-0.71})$ mas/yr is discrepant to this work $\sim 3\sigma$ in RA, and consistent to this work $\sim 2 \sigma$ in Dec (the measurements in Dec have larger uncertainties than in RA).

The direction of the lens-source relative proper motion $\varphi$ is not a property of the lens itself, but was an important point of comparison across previous work so we consider it here.
It is defined in \cite{Sahu:2022} as the position angle of the lens-source relative proper motion in equatorial coordinates.
In this work, we find $\varphi = 343.75^{+4.80}_{-3.95}$ deg, consistent with the measurements of \cite{Mroz:2022} $\varphi = 342.5 \pm 4.9$ deg and \cite{Sahu:2022} $\varphi = 342.3 \pm 3.0$ deg to $1 \sigma$.
The value of \cite{Lam:2022a,Lam:2022b} from the DW model $\varphi = 355.47^{+2.66}_{-2.11}$ is discrepant at $> 2\sigma$ from this work.

The photometric and astrometric measurements independently constrain $\varphi$.
Modeling the re-reduced OGLE photometry alone yields $\varphi = 345.1 \pm 3.7$ deg \citep{Mroz:2022}.
Modeling the re-reduced HST astrometry presented here alone measures $\varphi = 326.9^{+11.5}_{-11.0}$ deg.
The updated photometry and astrometry are consistent with each other to $<2\sigma$.

In general, the properties of the lens of OB110462 inferred in this work are somewhat discrepant with the \cite{Lam:2022a,Lam:2022b} DW model, and in reasonable agreement with \cite{Sahu:2022} and \cite{Mroz:2022}.
We also note that they are inconsistent with the \cite{Lam:2022a,Lam:2022b} EW model; the measurements of $M_L = 2.15^{+0.67}_{-0.54} M_\odot$, $D_L = 0.92^{+0.38}_{-0.22}$ kpc, $v_{T,L} = 7.26^{+4.88}_{-4.88}$ km/s, $(\mu_{L,E}, \mu_{L,N}) = (-0.69^{+0.91}_{-0.94}, 1.53^{+1.21}_{-1.12})$, and $\varphi = 18.08^{+8.60}_{-8.31}$ are discrepant $>3\sigma$ in mass, lens-source relative proper motion, and lens proper motion in RA and $> 4\sigma$ in transverse velocity with this work.

\begin{deluxetable}{lccc}
\tablecaption{Summary of data and models in different analyses
\label{tab:summary of analyses}}
\tablehead{
    \colhead{Reference} &
    \colhead{OGLE data} & 
    \colhead{HST data} & 
    \colhead{Likelihood}
    }
\startdata
\textbf{This work} & \textbf{Updated} & \textbf{2011-2022} & \textbf{DW}\\
Lam+22, DW & Original & 2011-2021 & DW\\
Lam+22, EW & Original & 2011-2021 & EW\\
Sahu+22 & Original & 2011-2017 & DW\\
Mroz+22 & Updated & 2011-2017 & DW
\enddata
\tablecomments{\emph{OGLE data}: ``Original" is the original OGLE reduction of OB110462 used in \cite{Lam:2022a,Lam:2022b,Sahu:2022}; ``Updated" is the OGLE re-reduction of OB110462 presented in \cite{Mroz:2022}.
\emph{HST data}: the timespan listed indicates the years of HST data used.
\emph{Likelihood}: ``DW" is the likelihood used in performing the model fit that weights each data point equally; ``EW" is the likelihood that weights the OGLE photometry, HST photometry, and HST astrometry equally.}
\end{deluxetable}

\begin{deluxetable*}{lcccccc}
\tablecaption{Comparison of lens properties
\label{tab:comparison}}
\tablehead{
    \colhead{Reference} &
    \colhead{$M_L (M_\odot)$} & 
    \colhead{$D_L$ (kpc)} & 
    \colhead{$v_{T,L}$ (km/s)} & 
    \colhead{$\mu_{L,E}$ (mas/yr)} &
    \colhead{$\mu_{L,N}$ (mas/yr)} &
    \colhead{$\varphi$ (deg)}
    }
\startdata
\textbf{This work} & $\mathbf{6.03^{+1.19}_{-1.04}}$ & $\mathbf{1.72^{+0.32}_{-0.23}}$ & $\mathbf{37.61^{+5.12}_{-5.13}}$ & $\mathbf{-3.80^{+0.48}_{-0.55}}$ & $\mathbf{2.60^{+0.83}_{-0.80}}$ & $\mathbf{343.75^{+4.80}_{-3.95}}$\\
Lam+22, DW & $3.79^{+0.62}_{-0.57}$ & $1.67^{+0.26}_{-0.20}$ & $23.95^{+2.95}_{-2.95}$ & $-2.64^{+0.18}_{-0.24}$ & $1.46^{+0.63}_{-0.71}$ & $355.47^{+2.66}_{-2.11}$\\
Lam+22, EW & $2.15^{+0.67}_{-0.54}$ & $0.92^{+0.38}_{-0.22}$ & $7.26^{+4.88}_{-4.88}$ & $-0.69^{+0.91}_{-0.94}$ & $1.53^{+1.21}_{-1.12}$ & $18.08^{+8.60}_{-8.31}$\\
Sahu+22 & $7.1 \pm 1.3$ & $1.58 \pm 0.18$ & $\sim 45 \pm 5.5$ & $-4.36 \pm 0.22$ & $3.06 \pm 0.66$ & $342.5 \pm 4.9$\\
Mroz+22 & $7.88 \pm 0.82$ & $1.62 \pm 0.15$ & $43.4 \pm 3.8$ & $-4.48 \pm 0.39$ & $3.29 \pm 0.5$ & $342.3 \pm 3.0$
\enddata
\tablecomments{Comparison of lens mass $M_L$, distance $D_L$, transverse velocity $v_{T,L}$, proper motion vector $(\mu_{L,E}$, $\mu_{L,N}$), and lens-source relative proper motion direction $\varphi$ inferred from various studies of OB110462.}
\end{deluxetable*}

\subsection{Goodness-of-fits \label{sec:Goodness-of-fits}}

\begin{figure*}
    \centering
    \includegraphics[width=1.0\linewidth]{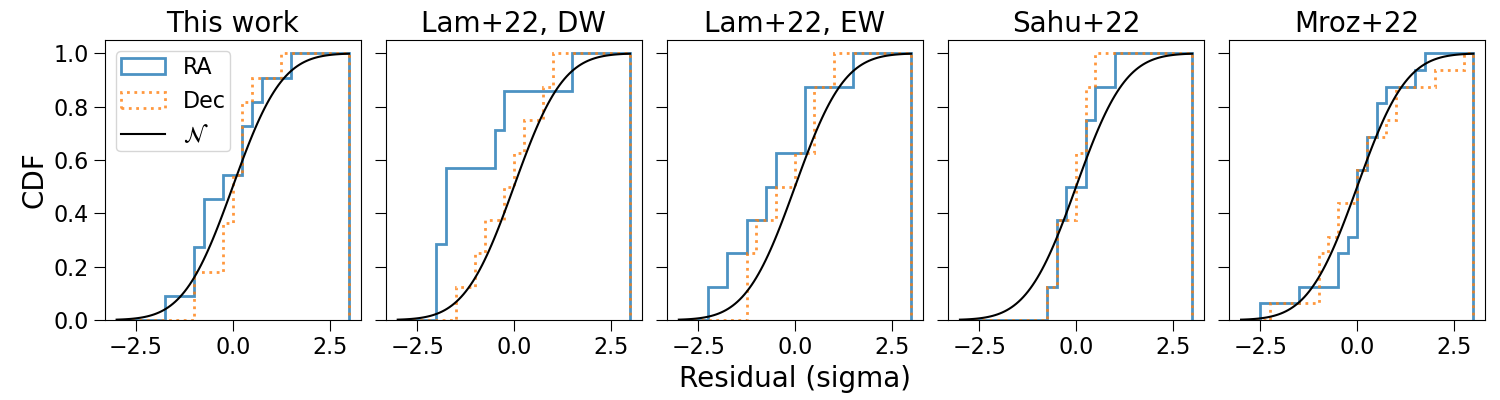}
    \caption{\label{fig:chi_xy_cdf}
    CDF of normalized residuals for astrometric models and data, as compared to the CDF of a standard normal distribution.
    The CDF of normalized residuals for the astrometry in RA (Dec) is shown in the blue solid (orange dotted) line.
    The CDF of a standard normal distribution is shown as the black curve.}
\end{figure*}

Next, we consider the goodness-of-fits of the astrometric model to the data.
As a reminder, the data and models used for each work are summarized in Table \ref{tab:summary of analyses}.
Figure \ref{fig:chi_xy_cdf} shows the CDF of the normalized residuals of the astrometric data and models against the CDF of a standard normal distribution.
For all the data sets except \cite{Sahu:2022}, the astrometry in F606W and F814W are separate data points; \cite{Sahu:2022} averages astrometry across both F606W and F814W filters to obtain a single position.
The residuals in this work are in good agreement with that of a standard normal, as are the residuals in \cite{Mroz:2022} and the \cite{Lam:2022a,Lam:2022b} EW model.
The residuals in \cite{Lam:2022a,Lam:2022b} are somewhat larger than expected in RA due to the model being a poor fit to the data.
The residuals in \cite{Sahu:2022} are smaller than expected, possibly indicative of underestimated uncertainties.

To be more quantitative, we also perform an Anderson-Darling (AD) test to check whether the distribution of normalized residuals is consistent with a standard normal distribution.
Table \ref{tab:fit_compare_ks_ad} lists the the AD test $S$-statistic, for the model fits in the RA and Dec.
The critical values for significance levels of 5\% and 1\% are 0.709 and 0.984, respectively.
Thus, all models presented are consistent with being drawn from a standard normal distribution.

\begin{deluxetable}{lcc}
\tablecaption{AD test statistic
\label{tab:fit_compare_ks_ad}}
\tablehead{
    \colhead{Work} &
    \colhead{RA} & 
    \colhead{Dec} 
    }
\startdata
\textbf{This work} & \textbf{0.39} & \textbf{0.24} \\ 
Lam+22, DW & 0.64 & 0.24 \\ 
Lam+22, EW & 0.17 & 0.27 \\ 
Sahu+22 & 0.22 & 0.40 \\ 
Mroz+22 & 0.50 & 0.29 \\ 

\enddata
\end{deluxetable}

\subsection{Bias correction method \label{sec:Bias correction method}}

\begin{figure*}[t!]
    \centering
    \includegraphics[width=1.0\linewidth]{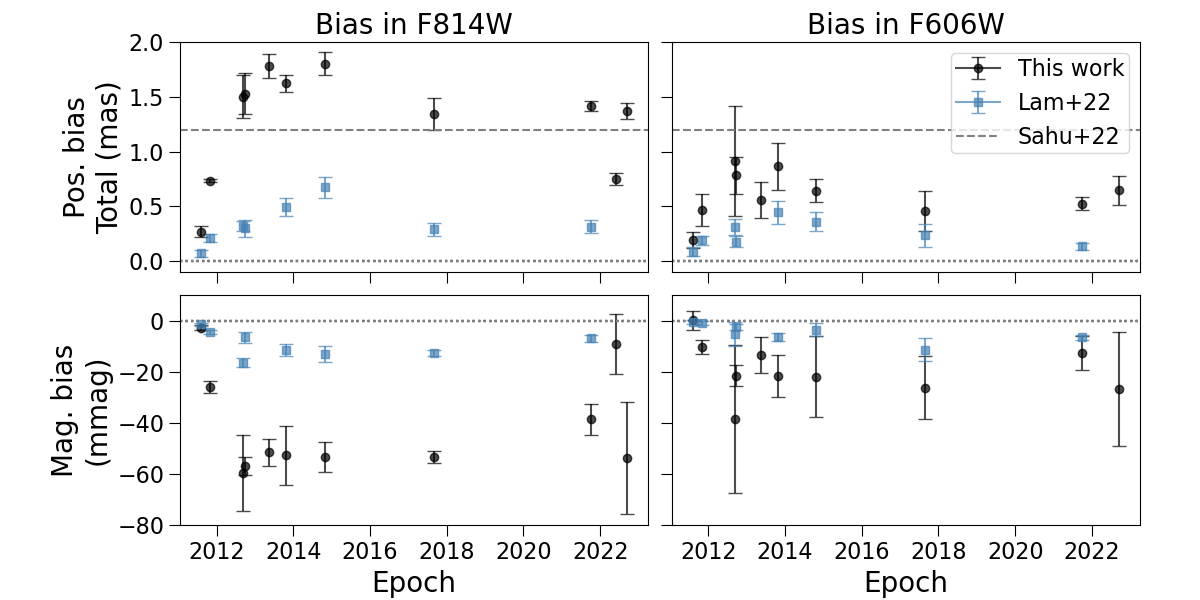}
    \caption{\label{fig:inject_recover_bias_comparison}
    Comparison of the bias correction measured in this work to those in \cite{Lam:2022b} and \cite{Sahu:2022}.
    The \emph{left (right) column} shows the bias in F814W (F606W).
    The \emph{top (bottom) row} shows the astrometric (photometric) bias, where the bias is defined as the recovered minus injected value.
    The bias measured in this work performing injection and recovery tests with \onepass\, is shown in \emph{black circles}.
    The bias measured in \cite{Lam:2022b} performing injection and recovery tests with \kstwo\, is shown as \emph{light blue squares}.
    The astrometric bias reported by \cite{Sahu:2022} performing PSF subtraction is shown as the \emph{dashed gray line}.
    \cite{Sahu:2022} report a single value for the astrometric bias; we show this same value in both the F814W and F606W panels.}
\end{figure*}

We find the main difference in the astrometry between this work and \cite{Lam:2022a,Lam:2022b} stems from the bias correction.
The other changes in the astrometric analysis did not significantly change the astrometry; we show this along with more detailed comparisons in Appendix \ref{app:Detailed comparison to previous work}.

We compare the measured photometric and astrometric bias corrections due to the bright neighbor star in Figure \ref{fig:inject_recover_bias_comparison}.
Using \onepass, we find the average positional bias in non-magnified epochs is around 1.6 mas in F814W and 0.6 mas in F606W (also see Figure \ref{fig:inject_recover_bias}, Table \ref{tab:injection recovery bias_eom}).
This is two to three times larger than the bias found by \cite{Lam:2022b} using \kstwo\, (c.f. Figures 22 and 23 in \cite{Lam:2022b}, the ``Neighbor-like" columns).
This suggests that PSF fitting in \kstwo\, is more precise than \onepass.

Similar to \cite{Lam:2022b}, there is minimal positional and magnitude bias in the first epoch, where OB110462 and the neighbor are of roughly equal brightness. 
In the third epoch onwards, the bias becomes non-negligible when OB110462 is much fainter, and the bias is primarily in the radial direction in F814W, and more mixed between radial and azimuthal in F606W.
We find that the magnitude bias to be also larger in \onepass\, than in \kstwo.
The average bias in non-magnified epochs is around 15 mmag in F814, and about 5 mmag in F606W when using \kstwo, as compared to around 50 mmag in F814W and 20 mmag in F606W when using \onepass. 
This again suggests that PSF fitting in \kstwo\, is more precise than \onepass.

The \onepass\, positional bias, when averaged across the two filters, is comparable to the bias of 1.2 mas\footnote{\cite{Sahu:2022} does not specify the bias as a function of filter; we assume that their stated bias of 1.2 mas is the average of F606W and F814W.} found by \cite{Sahu:2022}.
This is true, even though different sets of stars were used to compute the bias (18 vs. 4) as well as different methods (PSF subtraction vs. artificial star planting tests).

The results imply combining relative measurements across \kstwo\, and \onepass\, is not valid.
Although the majority of the underlying source extraction algorithms are identical, the particulars of PSF fitting are different enough to significantly alter the measured positions.
The two software source extraction methods cannot be combined together in a self-consistent manner.
Thus, the positional bias as calculated in \cite{Lam:2022a,Lam:2022b} was too small, and resulted in an incorrect set of astrometric measurements.

\begin{figure}
    \centering
    \includegraphics[width=\linewidth]{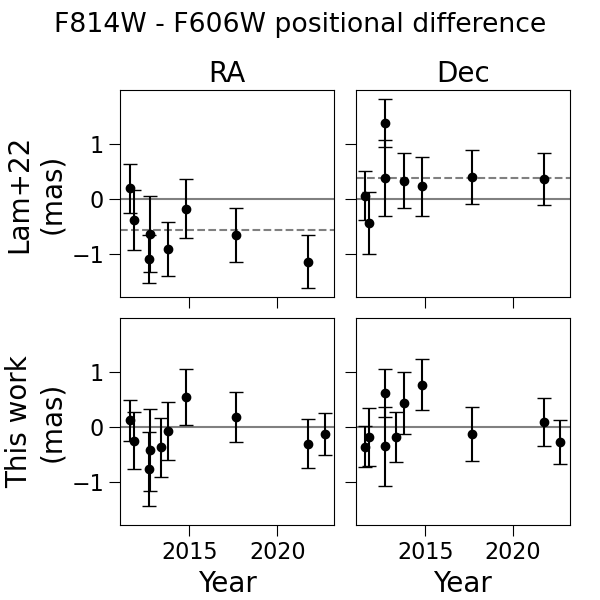}
    \caption{\label{fig:color_delta}
    Comparison of color differences in OB110462 astrometry.
    The \emph{top row} shows the difference in F606W and F814W position for OB110462 vs. time from \cite{Lam:2022a,Lam:2022b}.
    The bias correction calculated in \cite{Lam:2022b} has been applied, but the constant positional offset between F814W and F606W is \emph{not} included.
    The \emph{dashed line} is the ``empirical color offset" applied by \cite{Lam:2022b} to get the data in the two filters to better agree with each other.
    There is a significant difference between the positions measured between the two filters, especially in RA.
    The \emph{bottom row} shows the difference in F606W and F814W positions for OB110462 vs. time as derived in this work.
    The color difference has largely disappeared from the RA, and also slightly decreased in Dec.}
\end{figure}

In addition, \S 4.2.5 of \cite{Lam:2022b} noted an ``astrometric color offset" between the F814W and F606W positions of OB110462 and another microlensing event called OB110037.
For OB110462, the astrometry across the F814W and F606W filters were offset from each other by about 0.5 mas.
This color difference was tentatively attributed to binarity. 
For OB110462, this color difference can mainly be attributed to the bias correction (Figure \ref{fig:color_delta}).
However, the color difference in OB110037 and other stars are still unexplained.

\section{Discussion \label{sec:Discussion}}
\subsection{OB110462 in the context of the Galactic BH population \label{sec:OB110462 in the context of the Galactic BH population}}

\begin{figure*}
    \centering
    \includegraphics[width=1.0\linewidth]{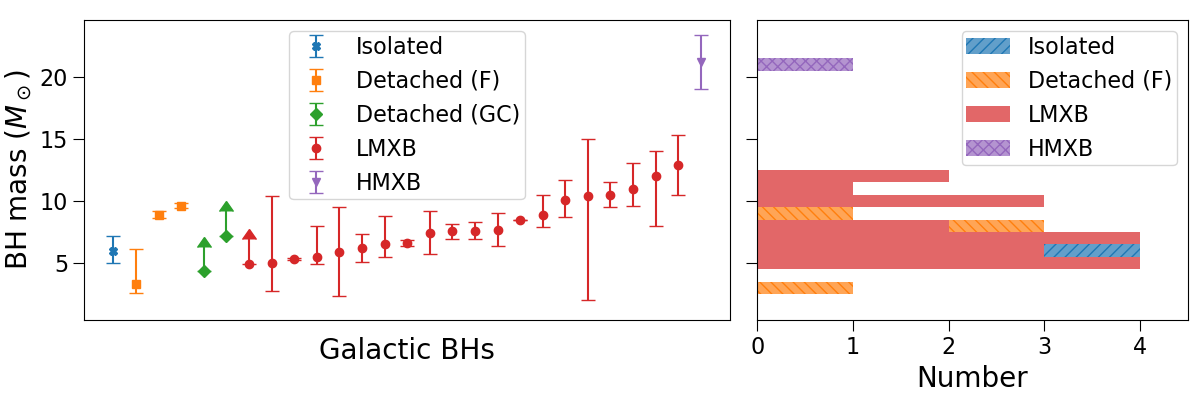} 
    \caption{\label{fig:galactic_bhs}
    OB110462 in the context of the observed Galactic BH population.
    \emph{Left:} Dynamically confirmed BHs, coded by their type (isolated, detached field binary, detached globular cluster binary, low-mass X-ray binary, high-mass X-ray binary).
    \emph{Right:} Histogram of Galactic BH mass measurements, coded by type.
    Note that not all BHs in the left panel are included in the right panel, as some only have lower limits on the mass.
    Note the histogram does not account for uncertainties on measurements.
    }
\end{figure*}

\begin{figure*}
    \centering
    \includegraphics[width=1.0\linewidth]{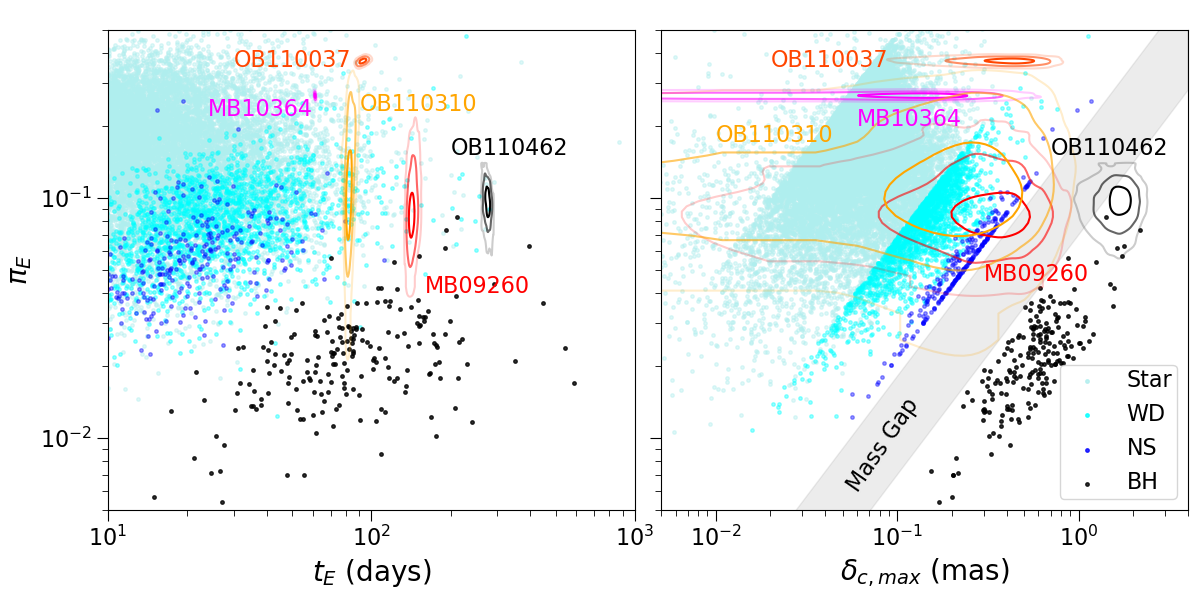} 
    \caption{\label{fig:microlensing_bhs}
    Microlensing parallax $\pi_E$ vs. Einstein crossing time $t_E$ (\emph{left}) and maximum astrometric shift $\delta_{c,max}$ (\emph{right}) of 5 BH microlensing candidates analyzed in \cite{Lam:2022a, Lam:2022b}.
    It is the same as Figure 6 of \cite{Lam:2022a}, except the contours for OB110462 have been updated with the values inferred in this work.
    MB09260, MB10364, OB110037, and OB110310 are low mass ($< 4 M_\odot$) lenses and not BHs.
    The points are simulated microlensing events from the \texttt{PopSyCLE} microlensing simulation \citep{Lam:2020}.
    Compared to the simulated BH population, OB110462 has a relatively large astrometric shift, large microlensing parallax, and long Einstein crossing time.
    }
\end{figure*}

Before 2019, all information about Galactic BHs came from X-ray binary systems, mainly low-mass X-ray binaries (LMXBs).
The observed population of BHs in LMXBs have masses tightly centered around $\sim 8 M_\odot$ \citep{Ozel:2010}.

A portrait of the complete Galactic BH population is finally emerging (Figure \ref{fig:galactic_bhs}; X-ray sources were compiled across catalogs from Aaron Geller (Northwestern)\footnote{\url{https://github.com/ageller/LIGO-Virgo-Mass-Plot_v2.0/blob/main/src/data/EMdata.json}},
Grzegorz Wiktorowicz and Chris Belczynski\footnote{\url{https://stellarcollapse.org/sites/default/files/table.pdf}},
and BlackCAT\footnote{\url{https://www.astro.puc.cl/BlackCAT/transients.php}} \citep{Corral-Santana:2016}).
Over the last 5 years, our knowledge of other types of Galactic BH systems has grown.
The first BH in a non-interacting system was found to have a somewhat surprising low mass of 3$M_\odot$, in the ``lower mass gap" where neutron stars and BHs had not previously been observed electromagnetically \citep{Thompson:2019}.
Since then, two more BHs in non-interacting binary systems have been found, with masses of $9 - 10 M_\odot$ \citep{El-Badry:2023_sun, Chakrabarti:2022, El-Badry:2023_rg}.
This is somewhat higher than the average observed BH mass in a LMXB, but falls within the typical mass range.

Now, OB110462 is the first isolated BH to have its mass measured.
At $6 M_\odot$, it is slightly lower than the average observed BH mass in an LMXB, although still falling within the typical mass range.

The selection effects that affect observations of BHs are important to consider if we are to understand the underlying population and BH mass function.
For example, observational selection effects may cause more massive BHs in LMXBs to be undetected \citep{Jonker:2021}.
For BHs in detached binaries, the picture is more muddled: there is tentative evidence of a dearth of BHs below $8 M_\odot$ \citep{El-Badry:2023_rg} as detected from Gaia, but ground-based RV surveys seem to not be finding these more massive BHs and have only found a single low-mass BH despite an observational bias toward higher masses \citep{Thompson:2019}.
For isolated BHs, there is also a selection bias towards more massive BHs as they have larger lensing cross sections; however, if there are many more low-mass BHs, those will dominate the observed sample.
With one detection, no strong conclusions can be made, but there should be at least as many $6-8 M_\odot$ BHs as $10 M_\odot$ BHs in the Galactic BH population.

Figure \ref{fig:microlensing_bhs} compares OB110462 to a simulated population of Galactic microlensing events.
OB110462 is somewhat unusual compared to typical microlensed BHs---it has a relatively large astrometric shift for its mass, as well as a large microlensing parallax for a BH.
This is due to OB110462 being a nearby ($<2$ kpc) lens; its large astrometric shift and microlensing parallax facilitated its detection and characterization.

Now that detections of BHs in various types of systems have been made, understanding their selection effects are needed to quantify their population properties.
This is the next research frontier that will enable us to understand the Galactic BH population as a whole.

\subsection{Origin and formation scenarios for OB110462}

With only a single isolated BH detection, tight constraints cannot be placed yet on their origins or specific formation scenarios.
In addition, OB110462 does not have full 6-D kinematic information available, since microlensing does not measure the lens radial velocity.
However, recent works examine possible situations that are consistent with observations of OB110462 and lay the groundwork for future studies.

Using statistical arguments and assuming OB110462 originated from a single star, \cite{Andrews:2022} finds that OB110462 is kinematically consistent with the Galactic thick disk.
Given the mass, distance, and transverse velocity found in this work, if OB110462 was born in the thick disk, natal kicks up to 100 km/s would be consistent with its current velocity.
On the other hand, if OB110462 formed in the thin disk and received a kick to a thick disk-like orbit, the kick would have had to be around 50-100 km/s.

\cite{Vigna-Gomez:2023} study the origins of isolated BHs.
They find that the majority of BHs with masses $< 10 M_\odot$ originated from binary systems, while the majority of BHs with masses $> 10 M_\odot$ originated as single stars.
This would imply that although OB110462 is now an isolated BH, it likely originated in a stellar binary system.

With additional mass measurements of isolated BHs, these studies and their extensions will be able to place increasingly tight constraints on the formation channels and origins of isolated BHs.
By performing targeted astrometric follow-up with existing facilities, it is possible to build the sample of isolated BHs to a few ($\lesssim 10$) over the next 5 to 10 years.

\subsection{Towards a large sample of isolated BHs}

\cite{Lam:2022a,Lam:2022b} analyzed a sample of 5 archival BH microlensing candidates, which included OB110462.
The other four candidates were not BHs.
They found that after accounting for selection effects, this single BH detection was consistent with $2 \times 10^8$ isolated Galactic BHs.
Although the results were consistent, they were not highly constraining due to the small sample size.
The Nancy Grace Roman Space Telescope (Roman), NASA's next flagship mission, presents the opportunity to find and characterize hundreds of isolated BHs with astrometric microlensing \citep{Lam:2020}, which will expand the sample size and enable stringent constraints on the Galactic BH population.

Roman will conduct several wide-field infrared surveys to answer questions about dark energy and dark matter, and find exoplanets \citep{Spergel:2015}.  
One of the surveys, the Galactic Bulge Time Domain Survey, nominally plans to observe $\sim 2$ deg$^2$ of the Galactic Bulge, finding several tens of thousands of microlensing events and a thousand exoplanets \citep{Penny:2019, Johnson:2020}. 
This survey also provides an excellent opportunity to find isolated BHs with microlensing.

OB110462, in addition to being a proof-of-concept of the method, raises several technical issues that have not been previously considered, and should be examined in preparation for using microlensing to find BHs with Roman.

Roman's mirror is the same size as Hubble (a diameter of 2.4 m), but will be observing at longer wavelengths than the optical bands used for this and most other microlensing work.
Thus the bias due to a nearby star in an event like OB110462 would be larger.

Roman will also be similarly or more undersampled than HST.
Roman's pixel scale is 110 mas/pix; the main filter for the Galactic Bulge Time Domain Survey is nominally F146, a wide filter centered at 1.46 micron corresponding to an angular resolution of $\theta_{res} = 153$ mas at the diffraction limit.
For comparison, the HST WFC3-UVIS pixel scale is 40 mas/pix; the resolution at the effective wavelength of F814W = 814 nm (F606W = 606 nm) is 85 (64) mas.
Thus, these PSF reconstruction and modeling methods currently required to achieve precise astrometry with HST will also be necessary for Roman.

In addition, the density of sources will be much higher in the infrared than in the (red) optical.
Roman will observe hundreds of millions of stars down to 25th magnitude in F146, exacerbating this issue.
In addition, if Roman chooses to observe fields closer to the Galactic plane towards $b=0$ deg, the density of sources will also increase. 
Finally, Roman's field of view will be 200 times that of Hubble's in the infrared.
Events like OB110462, where a faint source of interest is near a bright one, will be common and not necessarily an ignorable edge case.
A more automated or generalized way to correct the photometry and astrometry in this situation will be needed.

This also will have impact on the assumptions made in astrometry measurements.
This work as well as previous studies \citep{Lu:2016,Sahu:2022,Lam:2022a,Lam:2022b} assume that the astrometric shift measured is completely unblended; i.e. the lens is dark, there are no unrelated neighboring stars, and the only light that makes it to the telescope aperture is from the images of the source.
However, for all the reasons mentioned above, this assumption will not necessarily hold for Roman.
Despite the resolution gain from going to space, there is expected to be a non-neglible fraction of blended microlensing events \citep{Penny:2019}.
For dark lenses like BHs, the concern of blending would be from unrelated neighbor stars falling in Roman's aperture.
This extra flux would dilute the astrometric signal.
In addition, if the proper motion of the neighbor(s) was comparable to the lensing, it could affect not only the magnitude, but the shape of the astrometric shift.

Other considerations are how to best perform relative astrometry over such large fields of view.
The details of detector-to-detector calibration issues will be important.
Design tradeoffs in terms of observing strategy in order to attain sufficient astrometric precision should also be studied.
All the statements made here are qualitative, but these issues deserves further quantitative study.

\section{Conclusion \label{sec:Conclusion}}

We reanalyze OB110462, a microlensing event due to a dark, compact-object lens.
The astrometry we measure is significantly different from \cite{Lam:2022a,Lam:2022b}; the discrepancy is caused by a difference in the measured bias correction from a neighboring bright star.
By performing both the astrometric source extraction and bias correction measurement with the new version of \onepass\, in a self-consistent manner, we find our astrometry for OB110462 between 2011 and 2017 to be consistent with \cite{Sahu:2022}.
Modeling the updated HST photometry and astrometry along with the re-reduced OGLE photometry, we find OB110462 to be a BH with a mass of $6.03^{+1.19}_{-1.04} M_\odot$, consistent with the measurement of \cite{Sahu:2022}.
Thus, it appears so far that the masses of isolated Galactic BHs are similar to those in binary systems.
With the Roman Space Telescope, many more isolated BH systems can be characterized, and ultimately enable the measurement of the Galactic BH mass function.

\medskip

C.Y.L. and J.R.L. acknowledge support from the National Science Foundation under grant No. 1909641 and the Heising-Simons Foundation under grant No. 2022- 3542.
C.Y.L. also acknowledges support from NASA FINESST grant No. 80NSSC21K2043 and a research grant from the H2H8 Foundation.

We thank the anonymous referee and Dan Weisz for helpful comments that improved the manuscript.
We thank Kailash Sahu, Howard Bond, and Jay Anderson for proposing and taking the HST SNAP observation used in this work.

This research has made use of NASA’s Astrophysics Data System.
Some of the data presented in this paper were obtained from the Mikulski Archive for Space Telescopes (MAST) at the Space Telescope Science Institute. 
The specific observations analyzed can be accessed via \dataset[10.17909/gpw0-w659]{https://doi.org/10.17909/gpw0-w659}.

\appendix

\section{Comparison of pixel-based vs. tabular CTE correction \label{app:CTE}}

Here we compare the differences between the tabular and pixel-based CTE correction.
The differences alone do not indicate which type of correction is better or worse, but allows us to quantify the differences in the resultant astrometry.

\begin{figure*}
    \centering
    \includegraphics[width=1.0\linewidth]{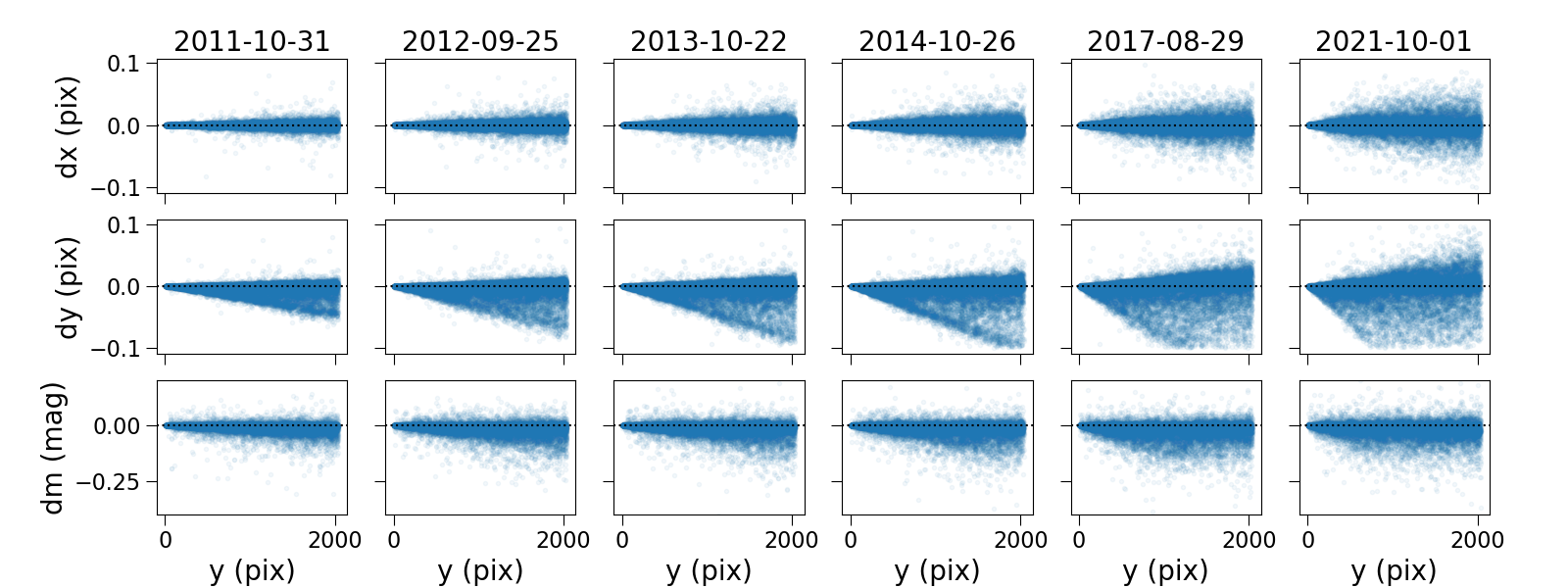} \\
    \includegraphics[width=1.0\linewidth]{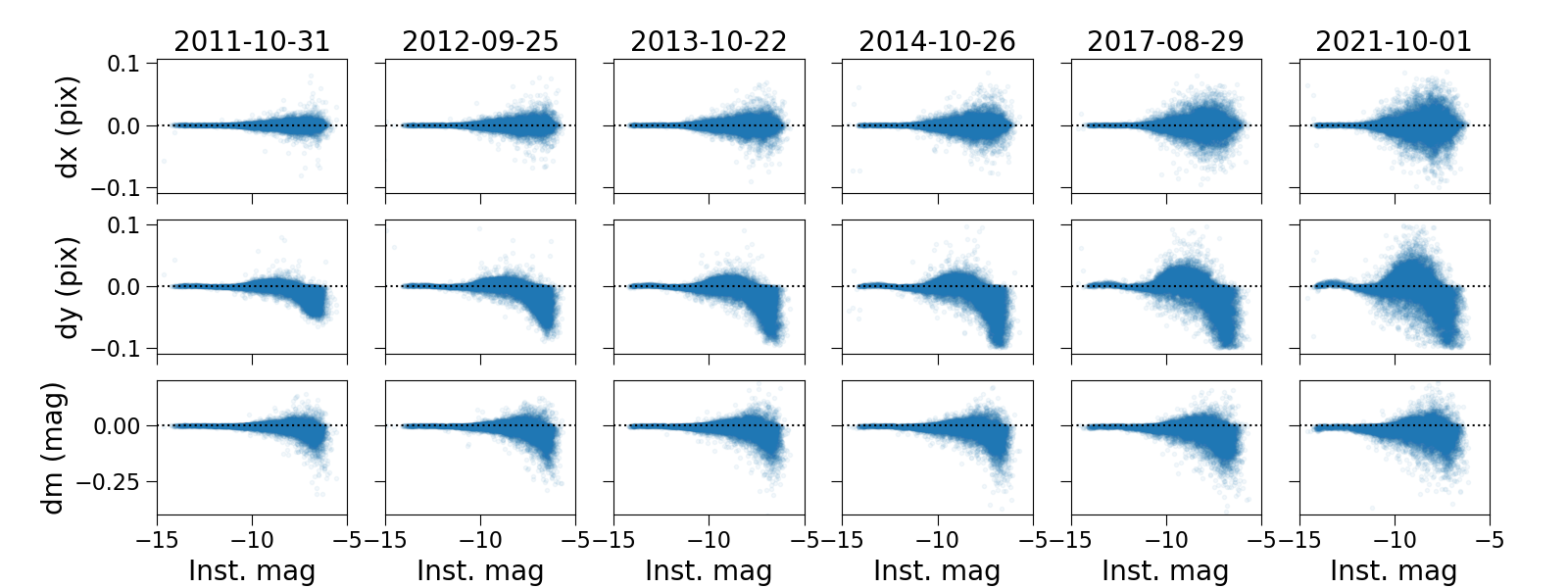}
    \caption{\label{fig:cte}
    Comparison of measured positions and magnitudes using new \onepass\, and tabular CTE correction vs. old \onepass\, and pixel-based CTE correction.
    These comparisons are performed on an single exposure, i.e. on the individual \texttt{flc.xym} or \texttt{flt.XYM} files. 
    The y-axis is the difference between the new tabular $-$ old 
    pixel based correction for $x$ detector position, $y$ detector position, and instrumental magnitude. 
    The top (bottom) plot shows the differences as a function of $y$ detector position (instrumental magnitude).
    }
\end{figure*}

Specifically, the comparison is made between the starlists in \cite{Lam:2022a,Lam:2022b} created using the pixel-based CTE correction (i.e. the old version of \onepass\, run on the $\texttt{flc}$ files) as compared to the starlists in this work created using the tabular CTE correction (i.e. the new version of \onepass, run on the $\texttt{flt}$ files).
Several frames spanning 10 years of the OB110462 dataset are shown in Figure \ref{fig:cte}.

As a function of $y$ detector position, there are only differences between the $y$ position and instrumental magnitude; the differences are symmetric in $x$ detector position.
This is expected, since CTE corrections only change the $y$ detector position and magnitude of the sources.
The scatter in x-position is likely due to minor differences between the versions of \onepass\, and differences in the choices of certain runtime parameters between the old and new reductions.
The scatter in $x$ position also allows us to see how much of the trend in $y$ position is due to scatter, vs. the CTE correction methods themselves.
As time goes on, the scatter increases greatly in the $x$ and $y$ position differences. 
The scatter also seems to increase with time for magnitude difference, although it is mostly constant after 2013 or so.

As a function of magnitude, the differences are again symmetric about $x$ position, but not for $y$ position and magnitude. 
In particular, for the differences in $y$ position before 2013, the bright stars with $m < -10$ are not affected, but after 2013 there is magnitude-dependent structure. 

\section{Priors \label{app:Priors}}

The priors for the fits are summarized in Table \ref{tab:priors}.
$\mathcal{N}(\mu, \sigma)$ denotes a normal distribution with mean $\mu$ and standard deviation $\sigma$. 
$\mathcal{N}_T(\mu, \sigma, l_\sigma, u_\sigma)$ denotes a normal distribution with a low end truncation at $\mu + \sigma l_\sigma$ and a high end truncation at $\mu + \sigma u_\sigma$.
$\mathcal{U}(a,b)$ denotes a uniform distribution from $a$ to $b$. 
$\Gamma^{-1} (\alpha, \beta)$ is the inverse gamma distribution (Equation G1 in \cite{Lam:2022b}).

The choice of priors is nearly identical to those in \cite{Lam:2022a,Lam:2022b}; see Appendix G of \cite{Lam:2022b} for details.
The only difference is for the photometric priors $b_{SFF}$ and $m_{base}$.
We changed the priors on $m_{base}$ as some of the photometry was recalibrated and the baseline magnitudes were now different.
We also made the priors on $b_{SFF}$ tighter; the blending is well constrained from previous work \citep{Lam:2022a,Sahu:2022,Mroz:2022}.
We did allow for some spread, but consider it well established that the OGLE lightcurve is highly blended ($b_{SFF}$ close to 0) and the HST lightcurves are less or unaffected by blending ($b_{SFF}$ close to 1).

\begin{deluxetable}{lc}
\tablecaption{Priors
\label{tab:priors}}
\tablehead{
    \colhead{Parameter} & 
    \colhead{Prior}
    }
\startdata
$t_0$ (MJD) & $\mathcal{N}(55770, 75)$\\
$u_0$ & $\mathcal{N}(0, 0.5)$\\
$t_E$ (days) & $\mathcal{N}_T(200, 100, -1.8, 3)$\\
$\pi_{E,E}$ & $\mathcal{N}(-0.02, 0.12)$\\
$\pi_{E,N}$ & $\mathcal{N}(-0.03, 0.13)$\\
$m_{base,O}$ (mag) & $\mathcal{N}(16.5, 0.1)$\\
$b_{SFF,O}$ & $\mathcal{U}(0, 0.2)$\\
$m_{base,H8}$ (mag) & $\mathcal{N}(19.9, 0.1)$\\
$b_{SFF,H8}$ & $\mathcal{U}(0.8, 1.1)$\\
$m_{base,H6}$ (mag) & $\mathcal{N}(22.0, 0.1)$\\
$b_{SFF,H6}$ & $\mathcal{U}(0.8, 1.05)$\\
\hline
$\log \sigma_{O}$ (mag) & $\mathcal{N}(0,5)$\\
$\rho_{{O}}$ (days) & $\Gamma^{-1}(0.53, 0.31))$\\
$\log \omega_{{0,O}}^4 S_{{0,O}}$ (mag$^2$ days$^{{-3}}$) & $\mathcal{N}(0.0001,5)$\\
$\log \omega_{0,O}$ (days$^{-1}$) & $\mathcal{N}(0,5)$\\
\hline
$\theta_E$ (mas) & $\mathcal{N}_T(-0.2, 0.3, -4, 4)$\\
$\pi_S$ (mas) & $\mathcal{N}_T(0.1126, 0.0213, -2.94, 90)$\\
$x_{{S0,E}}$ (arcsec) & $\mathcal{U}(0.229,0.233)$\\
$x_{{S0,N}}$ (arcsec & $\mathcal{U}(-0.240, -0.178)$\\
$\mu_{{S,E}}$ (mas/yr) & $\mathcal{U}(-2.472, 2.515)$\\
$\mu_{{S,N}}$ (mas/yr) & $\mathcal{U}(-2.354, 4.734)$\\

\enddata
\end{deluxetable}

\section{Detailed comparison to previous work \label{app:Detailed comparison to previous work}}

\subsection{CTE correction method}

First, we compare the pre-bias corrected astrometry presented in this work to that of \cite{Lam:2022a,Lam:2022b}.
The differences are the number of epochs of data used, the version of the \onepass\, software, the values of the additive error, and the method of CTE correction.
Of these differences, the method of CTE correction has the largest impact on the resultant astrometry.

We find no significant difference between the two astrometric time series after cross-epoch alignment and before implementing the bias correction.
While the different CTE corrections can change the measured positions of stars by several mas within a single epoch (Figure \ref{fig:cte}), these differences are effectively removed by the cross-epoch alignment when the starlists are transformed into a common reference frame using first and second-order polynomials.

Thus, the specific choice of CTE correction method, tabular or pixel-based, does not produce any significant change in the astrometric time series derived.
It is only important that CTE be corrected in some manner; in previous work we find that if no CTE correction of any kind is applied, there are systematics in the astrometry that cannot be removed by the first or second order polynomial transformations.

\subsection{Bias-corrected astrometric time series}

\begin{figure*}
    \centering
    \includegraphics[width=1.0\linewidth]{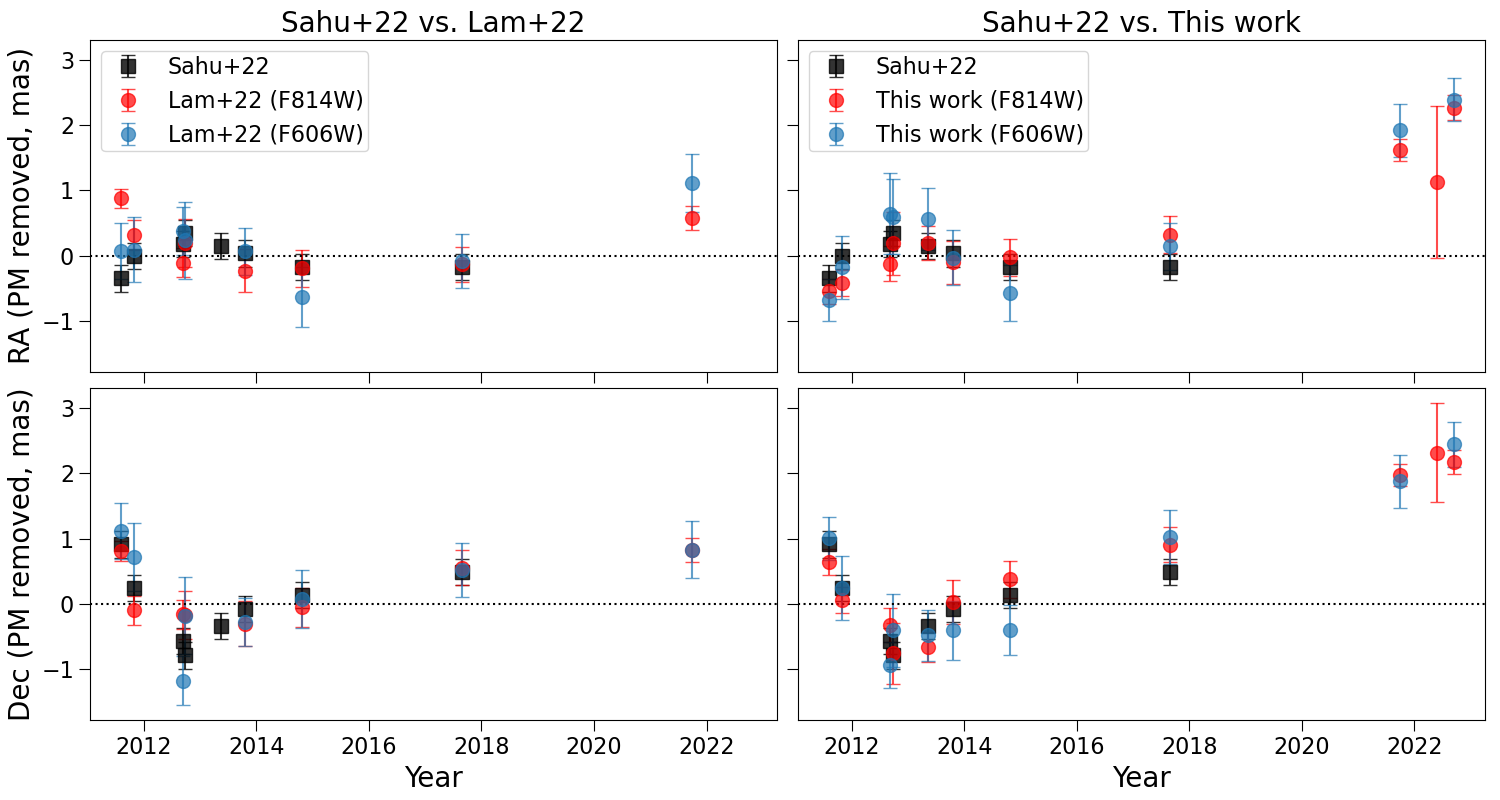}
    \caption{\label{fig:sahu_vs_lam_astrom}
    Comparison of HST astrometry derived by \cite{Sahu:2022}, \cite{Lam:2022a,Lam:2022b}, and this work.
    The \emph{top (bottom) row} shows the astrometry in RA (Dec) vs. time, with the source proper motion reported in \cite{Sahu:2022} (RA, Dec) = (-2.263, -3.597) mas/yr subtracted in order to more easily compare the astrometric deflections. 
    The \emph{left (right) column} compares the astrometry in \cite{Sahu:2022} to \cite{Lam:2022a,Lam:2022b} (this work).
    The positions of \cite{Sahu:2022} \emph{(black squares)} average all F814W and F606W measurements.
    \cite{Lam:2022a,Lam:2022b} and this work keep the F814W and F606W sets separate; F814W (F606W) positions are shown in red (blue) circles.
    }
\end{figure*}

Figure \ref{fig:sahu_vs_lam_astrom} directly compares the astrometry of OB110462 in \cite{Sahu:2022} (their Figures 16 and 18), \cite{Lam:2022a} (their Figure 2), and this work.
We emphasize that no microlensing model is being fit in Figure \ref{fig:sahu_vs_lam_astrom}.
Only a constant proper motion of (RA, Dec) = (-2.263, -3.597) mas/yr has been subtracted from the positions for easier visualization.
The subtracted proper motion was chosen to match the source proper motion inferred in \cite{Sahu:2022} for an ``apples-to-apples" comparison of the astrometry.

There is a clear discrepancy in the deflection in RA between the astrometry of \cite{Sahu:2022} and \cite{Lam:2022a,Lam:2022b} (\emph{top left panel}) across the magnified epochs (2011) vs. the non-magnified epochs (after 2011).
There is some minor discrepancy in Dec (\emph{top right panel}), but the measurements are generally within 1-2$\sigma$ of each other.
The new astrometry, including the updated bias correction, presented in this work is now consistent with the measurements of \cite{Sahu:2022} between 2011 and 2017.
In both RA (\emph{bottom left panel}) and Dec (\emph{bottom right panel}, the magnified and non-magnified astrometry are now all within $1\sigma$ of each other.

The new 2021-2022 measurements provide a more accurate source proper motion in baseline than in previous work. 
From Figure \ref{fig:sahu_vs_lam_astrom}, the source proper motion from our updated astrometric reductions appears to be different from that of \cite{Sahu:2022}.
For $t \gg t_0$, the slope of the proper motion-removed measurements should asymptotically approach 0. 
This is not the trend seen in the new 2021-2022 epochs, implying the source proper motion is different from the value inferred by \cite{Sahu:2022}.
Although the source proper motion is not particularly interesting in and of itself, its inferred value affects the measured astrometric shift, which in turn affects the measured lens mass.

We consider whether this different proper motion could be the result of systematics in the analysis of later epochs.
No systematic errors of 1-2 mas in the 2021-2022 astrometry were detected in the reference stars (Figures \ref{fig:by_eye_ref_stars1} and \ref{fig:by_eye_ref_stars2}).

The source proper motion inferred from the fit are presented in Table \ref{tab:DW fit}.
The uncertainties in the transformation from a relative astrometric reference frame to the absolute Gaia reference frame in which the proper motions are reported are 0.13 mas/yr and 0.11 mas/yr in RA and Dec, respectively.
\cite{Sahu:2022} do not state their uncertainties moving from their relative to absolute Gaia astrometric frame, but assuming their systematic uncertainties are comparable to ours, the source proper motion inferred in this work of (-2.02, -3.45) mas/yr and \cite{Sahu:2022} of (-2.263, -3.597) mas/yr are consistent within $1-2\sigma$.
Thus, the source proper motions across the model fits are consistent with each other.
However, there appears to be a hint of correlated residuals in the later epochs from 2014-2022 in RA (Figure \ref{fig:DW_astrometry}).

Additional observations in the future can establish the unlensed source proper motion and confirm or reject any potential discrepancy.

\subsection{Bias-corrected photometric time series}

\begin{figure}
    \centering
    \includegraphics[width=\linewidth]{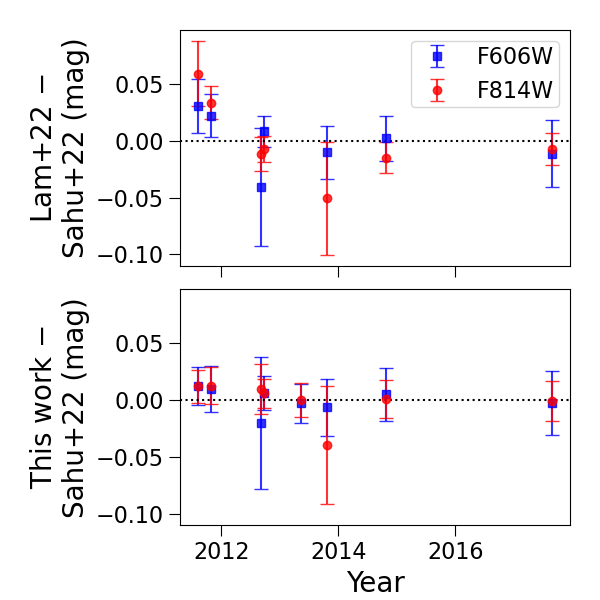} 
    \caption{\label{fig:sahu_vs_lam_phot_diff}
    Comparison of HST photometry derived by \cite{Sahu:2022}, \cite{Lam:2022a,Lam:2022b}, and this work.
    The \emph{top (bottom) row} compares the differences in the photometry in \cite{Sahu:2022} to \cite{Lam:2022a,Lam:2022b} (this work); red (blue) points denote data in the F814W (F606W) filter.
    }
\end{figure}

Figure \ref{fig:sahu_vs_lam_phot_diff} compares the HST photometry of OB110462 derived by \cite{Sahu:2022} (their Figure 13), \cite{Lam:2022b} (their Figures 7 and 8), and this work.
The calibration of the photometry across the three datasets slightly differs.
To calculate the difference between the zeropoint in the photometry of \cite{Lam:2022a} and this work as compared to \cite{Sahu:2022}, we calculate the average magnitude difference between those datasets and the \cite{Sahu:2022} photometry for the 7 common epochs between 2011 to 2017.
Then to put the photometry all onto comparable footing, this constant zeropoint difference is subtracted from the photometry in \cite{Lam:2022a} and this work.

There are minor differences between the photometry in \cite{Lam:2022a,Lam:2022b} and \cite{Sahu:2022} in the early (2011-2012) epochs.
With the updated bias correction in the photometry we derive in this work, we find the same photometry as measured by \cite{Sahu:2022} within the uncertainties.

Although the astrometric bias is the main reason for the updated mass result, Figure \ref{fig:sahu_vs_lam_phot_diff} illustrates the photometric bias should also not be neglected.
In particular, for Roman, with its extremely precise photometry, accounting for such systematics will be important.

\bibliography{sample631}{}

\begin{thebibliography}{}
\expandafter\ifx\csname natexlab\endcsname\relax\def\natexlab#1{#1}\fi
\providecommand{\url}[1]{\href{#1}{#1}}
\providecommand{\dodoi}[1]{doi:~\href{http://doi.org/#1}{\nolinkurl{#1}}}
\providecommand{\doeprint}[1]{\href{http://ascl.net/#1}{\nolinkurl{http://ascl.net/#1}}}
\providecommand{\doarXiv}[1]{\href{https://arxiv.org/abs/#1}{\nolinkurl{https://arxiv.org/abs/#1}}}

\bibitem[{{Anderson}(2014)}]{Anderson:2014}
{Anderson}, J. 2014, {The Impact of x-CTE in the WFC3/UVIS detector on
  Astrometry}, Instrument Science Report WFC3 2014-02, 9 pages

\bibitem[{{Anderson}(2021)}]{Anderson:2021_CTEtab}
---. 2021, {Table-Based CTE Corrections for flt-Format WFC3/UVIS}, Instrument
  Science Report WFC3 2021-13

\bibitem[{{Anderson}(2022)}]{Anderson:2022}
---. 2022, {One-Pass HST Photometry with hst1pass}, Instrument Science Report
  WFC3 2022-5, 55 pages

\bibitem[{{Anderson} {et~al.}(2021){Anderson}, {Baggett}, \&
  {Kuhn}}]{Anderson:2021_CTEv2.0}
{Anderson}, J., {Baggett}, S., \& {Kuhn}, B. 2021, {Updating the WFC3/UVIS CTE
  model and Mitigation Strategies}, Instrument Science Report 2021-9, 44 pages

\bibitem[{{Anderson} \& {King}(2006)}]{Anderson:2006}
{Anderson}, J., \& {King}, I.~R. 2006, {PSFs, Photometry, and Astronomy for the
  ACS/WFC}, Instrument Science Report ACS 2006-01

\bibitem[{{Anderson} {et~al.}(2008){Anderson}, {Sarajedini}, {Bedin}, {King},
  {Piotto}, {Reid}, {Siegel}, {Majewski}, {Paust}, {Aparicio}, {Milone},
  {Chaboyer}, \& {Rosenberg}}]{Anderson:2008}
{Anderson}, J., {Sarajedini}, A., {Bedin}, L.~R., {et~al.} 2008, \aj, 135,
  2055, \dodoi{10.1088/0004-6256/135/6/2055}

\bibitem[{{Andrews} \& {Kalogera}(2022)}]{Andrews:2022}
{Andrews}, J.~J., \& {Kalogera}, V. 2022, \apj, 930, 159,
  \dodoi{10.3847/1538-4357/ac66d6}

\bibitem[{{Bellini} {et~al.}(2018){Bellini}, {Libralato}, {Bedin}, {Milone},
  {van der Marel}, {Anderson}, {Apai}, {Burgasser}, {Marino}, \&
  {Rees}}]{Bellini:2018}
{Bellini}, A., {Libralato}, M., {Bedin}, L.~R., {et~al.} 2018, \apj, 853, 86,
  \dodoi{10.3847/1538-4357/aaa3ec}

\bibitem[{{Chakrabarti} {et~al.}(2022){Chakrabarti}, {Simon}, {Craig},
  {Reggiani}, {Guhathakurta}, {Dalba}, {Kirby}, {Chang}, {Hey}, {Savino}, \&
  {Geha}}]{Chakrabarti:2022}
{Chakrabarti}, S., {Simon}, J.~D., {Craig}, P.~A., {et~al.} 2022, arXiv
  e-prints, arXiv:2210.05003, \dodoi{10.48550/arXiv.2210.05003}

\bibitem[{{Corral-Santana} {et~al.}(2016){Corral-Santana}, {Casares},
  {Mu{\~n}oz-Darias}, {Bauer}, {Mart{\'\i}nez-Pais}, \&
  {Russell}}]{Corral-Santana:2016}
{Corral-Santana}, J.~M., {Casares}, J., {Mu{\~n}oz-Darias}, T., {et~al.} 2016,
  \aap, 587, A61, \dodoi{10.1051/0004-6361/201527130}

\bibitem[{{Dominik} \& {Sahu}(2000)}]{Dominik:2000}
{Dominik}, M., \& {Sahu}, K.~C. 2000, \apj, 534, 213, \dodoi{10.1086/308716}

\bibitem[{{El-Badry} {et~al.}(2023{\natexlab{a}}){El-Badry}, {Rix}, {Quataert},
  {Howard}, {Isaacson}, {Fuller}, {Hawkins}, {Breivik}, {Wong}, {Rodriguez},
  {Conroy}, {Shahaf}, {Mazeh}, {Arenou}, {Burdge}, {Bashi}, {Faigler}, {Weisz},
  {Seeburger}, {Almada Monter}, \& {Wojno}}]{El-Badry:2023_sun}
{El-Badry}, K., {Rix}, H.-W., {Quataert}, E., {et~al.} 2023{\natexlab{a}},
  \mnras, 518, 1057, \dodoi{10.1093/mnras/stac3140}

\bibitem[{{El-Badry} {et~al.}(2023{\natexlab{b}}){El-Badry}, {Rix}, {Cendes},
  {Rodriguez}, {Conroy}, {Quataert}, {Hawkins}, {Zari}, {Hobson}, {Breivik},
  {Rau}, {Berger}, {Shahaf}, {Seeburger}, {Burdge}, {Latham}, {Buchhave},
  {Bieryla}, {Bashi}, {Mazeh}, \& {Faigler}}]{El-Badry:2023_rg}
{El-Badry}, K., {Rix}, H.-W., {Cendes}, Y., {et~al.} 2023{\natexlab{b}}, arXiv
  e-prints, arXiv:2302.07880.
\newblock \doarXiv{2302.07880}

\bibitem[{{Fender} {et~al.}(2013){Fender}, {Maccarone}, \&
  {Heywood}}]{Fender:2013}
{Fender}, R.~P., {Maccarone}, T.~J., \& {Heywood}, I. 2013, \mnras, 430, 1538,
  \dodoi{10.1093/mnras/sts688}

\bibitem[{{Hog} {et~al.}(1995){Hog}, {Novikov}, \& {Polnarev}}]{Hog:1995}
{Hog}, E., {Novikov}, I.~D., \& {Polnarev}, A.~G. 1995, \aap, 294, 287

\bibitem[{{Johnson} {et~al.}(2020){Johnson}, {Penny}, {Gaudi}, {Kerins},
  {Rattenbury}, {Robin}, {Calchi Novati}, \& {Henderson}}]{Johnson:2020}
{Johnson}, S.~A., {Penny}, M., {Gaudi}, B.~S., {et~al.} 2020, \aj, 160, 123,
  \dodoi{10.3847/1538-3881/aba75b}

\bibitem[{{Jonker} {et~al.}(2021){Jonker}, {Kaur}, {Stone}, \&
  {Torres}}]{Jonker:2021}
{Jonker}, P.~G., {Kaur}, K., {Stone}, N., \& {Torres}, M. A.~P. 2021, \apj,
  921, 131, \dodoi{10.3847/1538-4357/ac2839}

\bibitem[{{Kuhn} \& {Anderson}(2021)}]{Kuhn:2021}
{Kuhn}, B., \& {Anderson}, J. 2021, {WFC3/UVIS: New FLC External CTE Monitoring
  2009-2020}, Instrument Science Report WFC3 2021-6, 15 pages

\bibitem[{{Lam} {et~al.}(2020){Lam}, {Lu}, {Hosek}, {Dawson}, \&
  {Golovich}}]{Lam:2020}
{Lam}, C.~Y., {Lu}, J.~R., {Hosek}, Matthew~W., J., {Dawson}, W.~A., \&
  {Golovich}, N.~R. 2020, \apj, 889, 31, \dodoi{10.3847/1538-4357/ab5fd3}

\bibitem[{{Lam} {et~al.}(2022{\natexlab{a}}){Lam}, {Lu}, {Udalski}, {Bond},
  {Bennett}, {Skowron}, {Mr{\'o}z}, {Poleski}, {Sumi}, {Szyma{\'n}ski},
  {Koz{\l}owski}, {Pietrukowicz}, {Soszy{\'n}ski}, {Ulaczyk}, {Wyrzykowski},
  {Miyazaki}, {Suzuki}, {Koshimoto}, {Rattenbury}, {Hosek}, {Abe}, {Barry},
  {Bhattacharya}, {Fukui}, {Fujii}, {Hirao}, {Itow}, {Kirikawa}, {Kondo},
  {Matsubara}, {Matsumoto}, {Muraki}, {Olmschenk}, {Ranc}, {Okamura}, {Satoh},
  {Silva}, {Toda}, {Tristram}, {Vandorou}, {Yama}, {Abrams}, {Agarwal}, {Rose},
  \& {Terry}}]{Lam:2022a}
{Lam}, C.~Y., {Lu}, J.~R., {Udalski}, A., {et~al.} 2022{\natexlab{a}}, \apjl,
  933, L23, \dodoi{10.3847/2041-8213/ac7442}

\bibitem[{{Lam} {et~al.}(2022{\natexlab{b}}){Lam}, {Lu}, {Udalski}, {Bond},
  {Bennett}, {Skowron}, {Mr{\'o}z}, {Poleski}, {Sumi}, {Szyma{\'n}ski},
  {Koz{\l}owski}, {Pietrukowicz}, {Soszy{\'n}ski}, {Ulaczyk}, {Wyrzykowski},
  {Miyazaki}, {Suzuki}, {Koshimoto}, {Rattenbury}, {Hosek}, {Abe}, {Barry},
  {Bhattacharya}, {Fukui}, {Fujii}, {Hirao}, {Itow}, {Kirikawa}, {Kondo},
  {Matsubara}, {Matsumoto}, {Muraki}, {Olmschenk}, {Ranc}, {Okamura}, {Satoh},
  {Silva}, {Toda}, {Tristram}, {Vandorou}, {Yama}, {Abrams}, {Agarwal}, {Rose},
  \& {Terry}}]{Lam:2022b}
---. 2022{\natexlab{b}}, \apjs, 260, 55, \dodoi{10.3847/1538-4365/ac7441}

\bibitem[{{Lu} {et~al.}(2016){Lu}, {Sinukoff}, {Ofek}, {Udalski}, \&
  {Kozlowski}}]{Lu:2016}
{Lu}, J.~R., {Sinukoff}, E., {Ofek}, E.~O., {Udalski}, A., \& {Kozlowski}, S.
  2016, \apj, 830, 41, \dodoi{10.3847/0004-637X/830/1/41}

\bibitem[{{Mereghetti} {et~al.}(2022){Mereghetti}, {Sidoli}, {Ponti}, \&
  {Treves}}]{Mereghetti:2022}
{Mereghetti}, S., {Sidoli}, L., {Ponti}, G., \& {Treves}, A. 2022, \apj, 934,
  62, \dodoi{10.3847/1538-4357/ac7965}

\bibitem[{{Miyamoto} \& {Yoshii}(1995)}]{Miyamoto:1995}
{Miyamoto}, M., \& {Yoshii}, Y. 1995, \aj, 110, 1427, \dodoi{10.1086/117616}

\bibitem[{{Mr{\'o}z} {et~al.}(2022){Mr{\'o}z}, {Udalski}, \&
  {Gould}}]{Mroz:2022}
{Mr{\'o}z}, P., {Udalski}, A., \& {Gould}, A. 2022, \apjl, 937, L24,
  \dodoi{10.3847/2041-8213/ac90bb}

\bibitem[{{Olejak} {et~al.}(2020){Olejak}, {Belczynski}, {Bulik}, \&
  {Sobolewska}}]{Olejak:2020}
{Olejak}, A., {Belczynski}, K., {Bulik}, T., \& {Sobolewska}, M. 2020, \aap,
  638, A94, \dodoi{10.1051/0004-6361/201936557}

\bibitem[{{{\"O}zel} {et~al.}(2010){{\"O}zel}, {Psaltis}, {Narayan}, \&
  {McClintock}}]{Ozel:2010}
{{\"O}zel}, F., {Psaltis}, D., {Narayan}, R., \& {McClintock}, J.~E. 2010,
  \apj, 725, 1918, \dodoi{10.1088/0004-637X/725/2/1918}

\bibitem[{{Penny} {et~al.}(2019){Penny}, {Gaudi}, {Kerins}, {Rattenbury},
  {Mao}, {Robin}, \& {Calchi Novati}}]{Penny:2019}
{Penny}, M.~T., {Gaudi}, B.~S., {Kerins}, E., {et~al.} 2019, \apjs, 241, 3,
  \dodoi{10.3847/1538-4365/aafb69}

\bibitem[{{Sabbi} {et~al.}(2016){Sabbi}, {Lennon}, {Anderson}, {Cignoni}, {van
  der Marel}, {Zaritsky}, {De Marchi}, {Panagia}, {Gouliermis}, {Grebel},
  {Gallagher}, {Smith}, {Sana}, {Aloisi}, {Tosi}, {Evans}, {Arab}, {Boyer}, {de
  Mink}, {Gordon}, {Koekemoer}, {Larsen}, {Ryon}, \& {Zeidler}}]{Sabbi:2016}
{Sabbi}, E., {Lennon}, D.~J., {Anderson}, J., {et~al.} 2016, \apjs, 222, 11,
  \dodoi{10.3847/0067-0049/222/1/11}

\bibitem[{{Sahu} {et~al.}(2022){Sahu}, {Anderson}, {Casertano}, {Bond},
  {Udalski}, {Dominik}, {Calamida}, {Bellini}, {Brown}, {Rejkuba}, {Bajaj},
  {Kains}, {Ferguson}, {Fryer}, {Yock}, {Mr{\'o}z}, {Koz{\l}owski},
  {Pietrukowicz}, {Poleski}, {Skowron}, {Soszy{\'n}ski}, {Szyma{\'n}ski},
  {Ulaczyk}, {Wyrzykowski}, {Barry}, {Bennett}, {Bond}, {Hirao}, {Silva},
  {Kondo}, {Koshimoto}, {Ranc}, {Rattenbury}, {Sumi}, {Suzuki}, {Tristram},
  {Vandorou}, {Beaulieu}, {Marquette}, {Cole}, {Fouqu{\'e}}, {Hill}, {Dieters},
  {Coutures}, {Dominis-Prester}, {Bennett}, {Bachelet}, {Menzies}, {Albrow},
  {Pollard}, {Gould}, {Yee}, {Allen}, {Almeida}, {Christie}, {Drummond},
  {Gal-Yam}, {Gorbikov}, {Jablonski}, {Lee}, {Maoz}, {Manulis}, {McCormick},
  {Natusch}, {Pogge}, {Shvartzvald}, {J{\o}rgensen}, {Alsubai}, {Andersen},
  {Bozza}, {Novati}, {Burgdorf}, {Hinse}, {Hundertmark}, {Husser}, {Kerins},
  {Longa-Pe{\~n}a}, {Mancini}, {Penny}, {Rahvar}, {Ricci}, {Sajadian},
  {Skottfelt}, {Snodgrass}, {Southworth}, {Tregloan-Reed}, {Wambsganss},
  {Wertz}, {Tsapras}, {Street}, {Bramich}, {Horne}, {Steele}, \& {RoboNet
  Collaboration}}]{Sahu:2022}
{Sahu}, K.~C., {Anderson}, J., {Casertano}, S., {et~al.} 2022, \apj, 933, 83,
  \dodoi{10.3847/1538-4357/ac739e}

\bibitem[{{Sana} {et~al.}(2012){Sana}, {de Mink}, {de Koter}, {Langer},
  {Evans}, {Gieles}, {Gosset}, {Izzard}, {Le Bouquin}, \&
  {Schneider}}]{Sana:2012}
{Sana}, H., {de Mink}, S.~E., {de Koter}, A., {et~al.} 2012, Science, 337, 444,
  \dodoi{10.1126/science.1223344}

\bibitem[{{Spergel} {et~al.}(2015){Spergel}, {Gehrels}, {Baltay}, {Bennett},
  {Breckinridge}, {Donahue}, {Dressler}, {Gaudi}, {Greene}, {Guyon}, {Hirata},
  {Kalirai}, {Kasdin}, {Macintosh}, {Moos}, {Perlmutter}, {Postman},
  {Rauscher}, {Rhodes}, {Wang}, {Weinberg}, {Benford}, {Hudson}, {Jeong},
  {Mellier}, {Traub}, {Yamada}, {Capak}, {Colbert}, {Masters}, {Penny},
  {Savransky}, {Stern}, {Zimmerman}, {Barry}, {Bartusek}, {Carpenter}, {Cheng},
  {Content}, {Dekens}, {Demers}, {Grady}, {Jackson}, {Kuan}, {Kruk}, {Melton},
  {Nemati}, {Parvin}, {Poberezhskiy}, {Peddie}, {Ruffa}, {Wallace}, {Whipple},
  {Wollack}, \& {Zhao}}]{Spergel:2015}
{Spergel}, D., {Gehrels}, N., {Baltay}, C., {et~al.} 2015, arXiv e-prints,
  arXiv:1503.03757.
\newblock \doarXiv{1503.03757}

\bibitem[{{Thompson} {et~al.}(2019){Thompson}, {Kochanek}, {Stanek}, {Badenes},
  {Post}, {Jayasinghe}, {Latham}, {Bieryla}, {Esquerdo}, {Berlind}, {Calkins},
  {Tayar}, {Lindegren}, {Johnson}, {Holoien}, {Auchettl}, \&
  {Covey}}]{Thompson:2019}
{Thompson}, T.~A., {Kochanek}, C.~S., {Stanek}, K.~Z., {et~al.} 2019, Science,
  366, 637, \dodoi{10.1126/science.aau4005}

\bibitem[{{Vigna-G{\'o}mez} \& {Ramirez-Ruiz}(2023)}]{Vigna-Gomez:2023}
{Vigna-G{\'o}mez}, A., \& {Ramirez-Ruiz}, E. 2023, \apjl, 946, L2,
  \dodoi{10.3847/2041-8213/acc076}

\bibitem[{{Walker}(1995)}]{Walker:1995}
{Walker}, M.~A. 1995, \apj, 453, 37, \dodoi{10.1086/176367}

\bibitem[{{Wiktorowicz} {et~al.}(2019){Wiktorowicz}, {Wyrzykowski},
  {Chruslinska}, {Klencki}, {Rybicki}, \& {Belczynski}}]{Wiktorowicz:2019}
{Wiktorowicz}, G., {Wyrzykowski}, {\L}., {Chruslinska}, M., {et~al.} 2019,
  \apj, 885, 1, \dodoi{10.3847/1538-4357/ab45e6}

\end{thebibliography}
\bibliographystyle{aasjournal}

\end{document}